\documentclass[12pt,preprint]{aastex7}

\usepackage{graphicx}

\newcommand{\SFR}{{\rm SFR}}

\newcommand{\magsec}{{\rm mag\, arcsec^{-2}}}

\shortauthors{Guo et al.}

\begin{document}

\title{Why Are Some Optically Red Spirals {\em NUV-r} Blue?}

\correspondingauthor{Cai-Na Hao}
\email{hcn@bao.ac.cn}

\author[0000-0002-3275-9914]{Rui Guo}
\affiliation{Tianjin Astrophysics Center, Tianjin Normal University, Tianjin 300387, China}
\email{glinka_welcome@163.com}

\author[0000-0002-0901-9328]{Cai-Na Hao}
\affiliation{Tianjin Astrophysics Center, Tianjin Normal University, Tianjin 300387, China}
\email{hcn@bao.ac.cn}

\author{Xiaoyang Xia}
\affiliation{Tianjin Astrophysics Center, Tianjin Normal University, Tianjin 300387, China}
\email{xyxia@bao.ac.cn}

\author[0000-0002-8614-6275]{Yong Shi}
\affiliation{School of Astronomy and Space Science, Nanjing University, Nanjing 210093, China}
\affiliation{Key Laboratory of Modern Astronomy and Astrophysics (Nanjing University), Ministry of Education, Nanjing 210093, China}
\email{yong@nju.edu.cn}

\author[0000-0002-9788-2577]{Lan Wang}
\affiliation{National Astronomical Observatory, Chinese Academy of Sciences, Datun Road 20A, Beijing 100012, China}
\affiliation{School of Astronomy and Space Science, University of Chinese Academy of Sciences, Beijing 100049, China}
\email{wanglan@bao.ac.cn}

\begin{abstract}

	To understand the complicated formation processes of disk galaxies, we
	carry out a comparative study for {\em NUV-r} blue and red spiral
	galaxies drawn from a parent sample of {\em u-r} red spirals with
	$M_{*} > 10^{10.5} M_{\odot}$ at $ 0.02 < z < 0.07$, based on the
	optical data from the Sloan Digital Sky Survey (SDSS) and the
	ultraviolet (UV) data from the Galaxy Evolution Explorer ({\it GALEX}).
	The analyses of the images and surface brightness profiles in the {\em
	NUV} and optical bands show that the differences between {\em NUV-r}
	blue and red spirals mainly occur in the outer disks (1-3 $R_{\rm e}$),
	and the contrast in {\em NUV} band is much larger than that in the
	optical bands. Both the positions on the star formation main sequence
	diagram and the {\em NUV-r} color profiles suggest that {\em NUV-r} red
	spirals have been fully quenched, whereas {\em NUV-r} blue spirals host
	quenched bulges and inner disks, as well as star-forming outer disks.
	Particularly, the disk mass-size relations indicate that, at a given
	disk mass, {\em NUV-r} blue spirals possess larger optical disks than
	{\em NUV-r} red spirals, by a factor of $\sim 1.20$. The environments
	and optical morphologies are consistent with the scenario that {\em
	NUV-r} blue spirals obtained fresh fuel for star formation either by
	interacting or merging with gas-rich galaxies or through accreting
	surrounding HI gas.

\end{abstract}


\section{INTRODUCTION\label{sec:intro}}

The formation and evolution of disk galaxies is an important subject in studies
of galaxies. In the local universe, disk galaxies constitute more than half of
the galaxy population \citep{Marzke1998,Conselice2005,Bamford2009}. Earlier
studies with the Hubble Space Telescope (HST) have revealed that the Hubble
Sequence was not yet established by $z\sim 1.5$, and irregular galaxies
dominate at $z > 2$ \citep[e.g.,][]{Conselice2005,Mortlock2013}. This
conclusion, however, could be biased due to the limited near-infrared (NIR)
coverage of HST (e.g., the F160W filter). These studies mostly sampled the
rest-frame UV or blue light for high-z ($z > 3$) galaxies, and hence tend to
emphasize irregular structures. Meanwhile, although the presence of $z\sim
1-3$ disk galaxies has been known for two decades \citep[e.g.,][]{Genzel2006,
ForsterSchreiber2006}, those studies mostly relied on gas kinematics and were
limited to small samples. The launch of the James Webb Space Telescope (JWST)
has altered this situation. With its superior sensitivity and spatial
resolution in the NIR and mid-infrared (MIR) wavebands, JWST has provided a
clear, direct mapping of the stellar mass distribution for statistically
significant samples of high-z galaxies. Incorporating the Atacama Large
Millimeter Array (ALMA) observations, large amounts of disk galaxies at $z > 2$
were discovered, and some of them are even quenched
\citep{Ferreira2022,Fudamoto2022,Cheng2023,Ferreira2023,
Jacobs2023,Kartaltepe2023, Nelson2023,Roman-Oliveira2023,Sun2024,Weibel2024}. 

It was known that late-type galaxies distribute differently from
early-type galaxies in the color-magnitude or color-stellar mass diagrams
\citep[e.g.,][]{Strateva2001, Schawinski2014}. The former mainly occupy the
star-forming blue cloud, whereas the latter are predominantly located in the
red sequence. This suggests that disk galaxies are star-forming, and the
quenching of star formation is generally accompanied by a morphological
transformation from disk galaxies to ellipticals. Nonetheless, this point of
view does not apply to red passive spiral galaxies
\citep[e.g.,][]{vandenBergh1976, Lee2008, Dressler1999, Poggianti1999,
Goto2003, Skibba2009, Bundy2010, Masters2010, FraserMcKelvie2016,
FraserMcKelvie2018}. Recently, \citet{FraserMcKelvie2022} carefully analyzed a
volume-limited sample of nearby galaxies drawn from the full sample of the
Mapping Nearby Galaxies at Apache Point Observatory
\citep[MaNGA,][]{Bundy2015}, and they found that passive disk galaxies
represent about 30\% of quenched population in both number and stellar mass in
the local universe. In addition, by investigating the kinematic properties of
1,862 galaxies based on MaNGA survey, \citet{Brownson2022} claimed that it is a
misleading viewpoint that disk galaxies are star-forming and spheroids are
quenched. \citet{Cortese2022} also pointed out that the structures of passive
galaxies are more heterogeneous than those of star-forming galaxies, based on
the studies of the correlation between the central stellar surface density
$\Sigma_1$ and the stellar spin parameter $\lambda_{re}$ for galaxies on and
below the star formation main sequence (SFMS).

In the literature, the selection criteria for passive spiral galaxies vary.
Although the detailed criteria can be very different, they fall into two broad
categories. One is based on colors, and the other relies on their positions on
the SFMS diagram or the specific star formation rate (sSFR). Given the easier
availability of optical imaging observations than those in other wavebands,
optical colors were often used to select red passive spirals. However, as
already pointed out by some papers \citep[e.g.,][]{Cortese2012,
FraserMcKelvie2016}, optically red spiral galaxies can include contaminants
that are either heavily dust obscured systems or objects with residual
low-level star formation. For the dust-related contamination, the exclusion of
edge-on galaxies often helps. Furthermore, IR color provides a widely used
means to distinguish genuinely passive systems from dusty star-forming galaxies
\citep[e.g.,][]{FraserMcKelvie2016, Mahajan2020, Pak2021}. To mitigate the
contamination from galaxies with low-level star formation, the near-ultraviolet
(NUV) waveband in combination with an optical band is a good choice,
considering that {\em NUV} is more sensitive to recent star formation than even
the bluest optical {\em u} band \citep{Wyder2007}. As for the criteria based on
the sSFR, aperture bias is an important issue \citep{Cortese2020}, which can be
alleviated by employing better measurements of global SFRs. In spite of these
improvements, galaxy samples selected by sSFR are sometimes inconsistent with
those selected via {\em NUV-r} colors \citep[e.g.,][]{Lemonias2014}. Therefore,
we must be careful to understand the conclusions drawn upon galaxy samples with
different selection criteria.

To utilize the large datasets provided by the Sloan Digital Sky Survey
\citep[SDSS,][]{York2000}, we employed the optical {\em u-r} colors to
construct samples of nearby massive ($M_{*} > 10^{10.5} M_{\odot}$) red spiral
galaxies, red ellipticals and blue spirals, and carried out a series of
comparative studies \citep{Hao2019, Guo2020, Zhou2021}. As the bluest optical
broadband filter, $u$-band is a better tracer of young stellar populations than
those at longer wavelengths. We therefore expect to obtain a cleaner sample of
passive galaxies by employing the {\em u-r} colors in place of the {\em g-r}
colors. The contamination from galaxies with low-level star formation can still
be an issue though. Our studies found that the formation epoch and star
formation timescales of red spirals are similar to elliptical galaxies with
similar stellar mass. Specifically, the star formation history analysis
revealed that more than 50\% and 90\% of their present day stellar mass formed
before $\sim 10$ and $\sim 6$ Gyrs ago, respectively \citep{Zhou2021}.
Furthermore, it was found that about half of the optically red spirals have HI
detections \citep{Guo2020, Wang2022}, suggesting possible contamination
from galaxies with residual star formation.

Samples of passive disk galaxies that were found to host a large amount of
HI also include non-genuine passive galaxies. \citet{Lemonias2014} carried
out high resolution HI observations using the Very Large Array (VLA) for 20
massive disk galaxies with rich HI content but low sSFR. They found that all
their sample galaxies have very extended HI disks, with radii of several tens
kpc and low HI surface densities. However, as indicated in their paper, the
galaxies with suppressed sSFR do not belong to the red sequence in the {\em
NUV-r} vs.  $M_*$ diagram. In addition, based on the Arecibo Legacy Fast ALFA
(ALFALFA), {\it Galaxy Evolution Explorer} Arecibo SDSS Survey (GASS) and COLD
GASS survey databases, \citet{Zhang2019} claimed that the vast majority of
massive quiescent central disk galaxies have a large amount of HI gas located
in regularly rotating disks. This claim was soon questioned by
\citet{Cortese2020}, who pointed out that the SDSS fiber-based
aperture-corrected SFR adopted by \citet{Zhang2019} is not a fair
representation of the global SFR. Actually, the SDSS images of quiescent disk
galaxies shown in \citet{Zhang2019} have already exhibited blue colors in the
outer regions. \citet{Parkash2019} studied a sample of 91 HI galaxies with
little or no star formation, and found that at least 32 out of 62 galaxies with
deep optical images show low levels of star formation in the outer regions. By
employing the combination of NUV from the {\it Galaxy Evolution Explorer} ({\it
GALEX}) and mid-infrared (MIR) photometry
from the Wide-field Infrared Survey Explorer (WISE) as a SFR estimator,
\citet{Cortese2020} concluded that the majority of nearby passive disks in the
extended GALEX Arecibo SDSS Survey (xGASS) sample lack HI gas reservoirs.

Nonetheless, these galaxies with residual star formation may represent a
special evolutionary phase. It has been recognized that both star formation
and quenching processes are complicated and diverse. For example,
\citet{Tacchella2022} investigated the star formation history of 161 massive
galaxies at redshift $\sim 0.8$. They found that star-forming galaxies evolve
along the SFMS, but cross the SFMS ridgeline several times in their formation
history. Such an oscillation around the SFMS has been envisioned as a
consequence of alternate gas depletion and accretion \citep{Tacchella2016,
Wang2019}. For quiescent galaxies, the quenching timescales and epochs exhibit
wide and continuous distributions. The authors also identified a small fraction
of galaxies (9/161) with significant rejuvenation. Even in the local universe,
some star-forming galaxies could be rejuvenated systems
\citep[e.g.,][]{Rathore2022,Hao2024,Tanaka2024}.

In this paper, we will employ NUV emission, a tracer of recent star
formation, to compare the properties of optically-selected red spirals with
different {\em NUV-r} colors, particularly focusing on the distributions of
their NUV emission, mass-size relations and environments, to understand the
evolutionary stage of {\em NUV-r} blue but optically red spiral galaxies. In
Section 2, we describe the sample selection and parameter derivation.
The results are presented in Section 3. In Sections 4 and 5, we
discuss and summarize our findings, respectively. Throughout this paper, we adopt the
\citet{Chabrier2003} initial mass function (IMF) and a cosmology with $H_{\rm
0}=70\,{\rm km \, s^{-1} Mpc^{-1}}$, $\Omega_{\rm m}=0.3$ and $\Omega_{\rm
\Lambda}=0.7$.

\section{SAMPLE AND PARAMETER}
\subsection{Sample Selection\label{sec:sample}}

As one of our series of work on {\em u-r} selected red spirals, the procedure
to obtain the optically red spiral sample here is similar to that of
\citet{Guo2020}, but it extends to slightly higher redshifts to include more
galaxies with {\em NUV} observations from {\it GALEX}.
In brief, we first selected galaxies with $0.02<z<0.07$ and $M_{\ast}
> 10^{10.5} M_{\odot}$ from the mass catalog of \citet{Mendel2014}, which
produced a sample of 37,177 galaxies. Then we picked out 12,688 spiral galaxies
from this sample, based on the morphological classification in Galaxy Zoo 1
\footnote{http://data.galaxyzoo.org/} \citep[GZ 1,][]{Lintott2008,
Lintott2011}. Spiral galaxies with an axis ratio $b/a < 0.5$ were further
removed to minimize the dust effect on color measurements. This yielded 6,156
face-on spiral galaxies.
Among these, we classified 839 as optically red,
by applying the criterion $ u-r > -0.073+0.227 \log (M_*/M_{\odot}) $ to the
diagram of dust-corrected {\em u-r} colors (from SDSS model magnitudes) as a
function of stellar mass.

The SDSS model magnitudes in all five ({\em u, g, r, i, z}) bands were
derived based on the best-fit model profile (either a de Vaucouleurs bulge
or an exponential disk) determined in the {\em r} band, with only the
normalization allowed to vary in the other bands \citep{Stoughton2002,
Abazajian2004}. Although they were measured in the same apertures, for
galaxies with a mixture of bulge and disk components, the large color gradients
can cause the integrated colors to deviate from their true values
\citep{Simard2011,Meert2016}. To obtain a more reliable color-selected sample,
we carried out matched-aperture photometry on SDSS {\em u, r} and {\it GALEX}
NUV bands images, which were retrieved from the NASA Sloan Atlas
(NSA)\footnote{http://www.nsatlas.org/} \citep[]{Blanton2011}.  The SDSS {\em
u-} and {\em r-}band images have similar point spread function (PSF) sizes,
whereas the PSFs in the {\it GALEX} NUV images are much broader.  Therefore, we
performed the photometry directly for the calculation of the {\em u-r} colors,
whereas we matched the PSF sizes first for the derivation of the {\em NUV-r}
colors. Elliptical apertures were used in the photometry. The ellipticities and
azimuth angles were extracted from the NSA catalogue, and the semi-major axes
of the elliptical apertures were determined by the 25 mag\,arcsec$^{-2}$
isophote in the SDSS {\em r} band. Finally, we corrected for the foreground
Galactic extinction provided by the NSA catalog, which originates from
\citet{Schlegel1998}.

\begin{figure}
\plotone{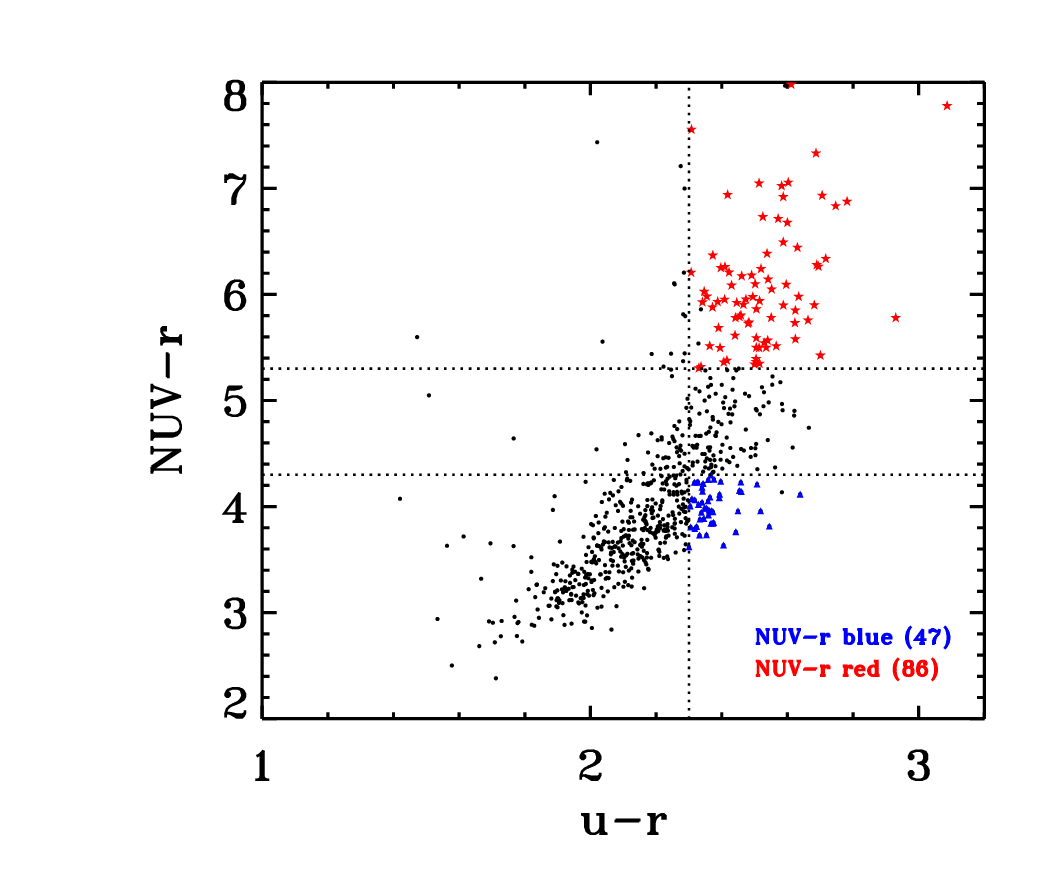}
\caption{
Matched aperture {\em NUV-r} color vs. {\em u-r} color diagram for massive
red spirals selected by the {\em u-r} colors derived from SDSS model magnitudes.
Blue triangles and red stars represent the final {\em NUV-r} blue and red
sample spirals.
The vertical dotted line represents $u-r=2.3$. The horizontal dotted lines
represent $NUV-r=4.3$ and $NUV-r=5.3$, respectively.
}
\label{redsample.ps}
\end{figure}

Among the 839 red spirals selected using the dust-corrected {\em u-} and {\em
r-} bands model magnitudes, there are 823 included in the NSA. A further
requirement of the availability of the {\em NUV} images reduced the sample size
to 707 galaxies. The distribution of these 707 spiral galaxies in the {\em
NUV-r} versus {\em u-r} diagram is shown in Figure \ref{redsample.ps}.  It is
obvious that the red spirals selected by their SDSS model {\em u-r} colors span
a wide range in matched-aperture {\em u-r} color. This suggests that the u-band
model magnitudes miss the fluxes in the outer parts of galaxies. As we shall
see in Section \ref{subsec:SBP}, the SDSS model colors mainly represent the
colors of our sample galaxies within $\sim 1\,Re$, instead of the entire
galaxy. In other words, our initial sample of red spirals includes a large
fraction of galaxies lying in the green valley and even some
residing in the blue cloud. Out of the 707 (823) galaxies, there are only
245 (292) with $u-r > 2.3$. However, we examined the results in \citet{Guo2020}
using the refined sample by excluding galaxies with $u-r \le 2.3$, and found
that the conclusions remain the same except for the HI detection rate. In
the refined sample, 28\% of red spirals have HI detections, compared to 45\% in
\citet{Guo2020}.

We further separated the 245 red spirals with $u-r > 2.3$ into {\em NUV-r} blue
and red galaxies, using $NUV-r < 4.3$ and $NUV-r > 5.3$, respectively. The
color threshold of $NUV-r > 5.3$ was determined according to the color
distributions of the passive spirals in \citet{FraserMcKelvie2016}, which is
slightly different from the commonly used $NUV-r > 5$ in the literature
\citep[e.g.,][]{Salim2014,Li2024}. Meanwhile, to enlarge the sample size
of {\em NUV-r} blue spirals, we changed the color cut from the widely used
$NUV-r < 4$ to $NUV-r < 4.3$. After removing galaxies contaminated by bright
stars, this selection yielded a final sample of 47 {\em NUV-r} blue and 86 {\em
NUV-r} red galaxies. Their basic information and Galactic-extinction-corrected
colors are listed in Table \ref{table1}. We note that the quantitative
results slightly depend on the specific color thresholds adopted, whereas the
qualitative conclusions are robust.

\startlongtable
\begin{deluxetable}{ccccccc}
\tabletypesize{\footnotesize}
\tablecolumns{7}
\tablewidth{0pt}
\tablecaption{Colors and environment of the {\em NUV-r} blue and red spirals}
\tablehead{
\colhead{ObjID} & \colhead{RA (J2000)} & \colhead{Dec (J2000)} & \colhead{Redshift} &
\colhead{{\em u-r}} & \colhead{{\em NUV-r}} &
\colhead{Environment} \\
\colhead{ } & \colhead{(deg)} & \colhead{(deg)} & \colhead{ } &
\colhead{(mag)} & \colhead{(mag)} & \colhead{} \\
\colhead{(1)} & \colhead{(2)} & \colhead{(3)} & \colhead{(4)} &
\colhead{(5)} & \colhead{(6)} & \colhead{(7)}}
\startdata
\cutinhead{{\em NUV-r} blue spirals}
587732484360044639 & 183.79126 & 50.68529 & 0.02974 & 2.38 & 4.26 & satellite \\
587733429770977288 & 222.55058 & 55.11942 & 0.04440 & 2.35 & 3.98 & isolated \\
587733609085337761 & 235.79839 & 47.75365 & 0.03821 & 2.46 & 4.23 & central \\
587730775484268958 & 324.51025 & 12.53612 & 0.04553 & 2.54 & 3.81 & isolated \\
587731868024111184 & 182.45949 & 54.00175 & 0.05024 & 2.36 & 4.05 & central \\
587735343728033922 & 149.14658 & 10.09465 & 0.05183 & 2.34 & 4.04 & central \\
587735742616961341 & 241.77399 & 30.21762 & 0.05469 & 2.37 & 3.84 & central \\
587736543098568872 & 230.39474 &  7.57642 & 0.04756 & 2.33 & 4.02 & satellite \\
587736914605768984 & 221.82343 & 11.13016 & 0.05354 & 2.36 & 3.92 & satellite \\
587737826209890568 & 116.09793 & 44.52774 & 0.04968 & 2.37 & 3.96 & isolated \\
587742012741582989 & 171.57805 & 21.09623 & 0.04285 & 2.32 & 4.23 & central \\
587730818433286308 & 331.79794 & -7.49541 & 0.06209 & 2.32 & 3.81 & satellite \\
587731499185799239 & 155.49457 & 53.01943 & 0.06255 & 2.30 & 3.62 & central \\
588011219682394142 & 240.42902 & 48.82793 & 0.04287 & 2.37 & 3.85 & satellite \\
588017703484326010 & 219.50909 & 10.16236 & 0.02794 & 2.30 & 4.00 & central \\
587741603103965367 & 180.18446 & 28.32862 & 0.05040 & 2.45 & 3.96 & central \\
587725073916821776 & 135.80276 & -0.75293 & 0.04834 & 2.31 & 4.07 & isolated \\
587735240640102602 & 136.37048 & 33.38861 & 0.06499 & 2.31 & 3.81 & satellite \\
587735241713451178 & 134.83943 & 33.64342 & 0.05951 & 2.33 & 3.73 & isolated \\
587736941994705213 & 245.50871 & 22.78210 & 0.06265 & 2.37 & 4.30 & isolated \\
587728932418945227 & 140.12738 & 51.67600 & 0.06552 & 2.37 & 3.95 & isolated \\
587737808494985297 & 120.10622 & 47.93469 & 0.05548 & 2.33 & 3.88 & isolated \\
588013382726189329 & 130.14026 & 36.59297 & 0.05373 & 2.37 & 3.84 & isolated \\
587739405714260127 & 200.91531 & 32.75106 & 0.06091 & 2.36 & 4.09 & isolated \\
587739458836693043 & 223.38120 & 26.96920 & 0.06113 & 2.36 & 4.25 & central \\
587739648348455067 & 148.68144 & 27.91859 & 0.05600 & 2.32 & 4.06 & isolated \\
587739809417986058 & 224.99933 & 20.77002 & 0.06159 & 2.32 & 4.23 & isolated \\
588017992306327833 & 218.29836 &  9.20888 & 0.05358 & 2.41 & 3.64 & central \\
587741491446481037 & 170.42123 & 30.19954 & 0.05843 & 2.35 & 3.89 & isolated \\
587741534402183260 & 186.71584 & 31.09422 & 0.05981 & 2.45 & 4.15 & satellite \\
587725470666850580 & 123.80689 & 46.72039 & 0.05539 & 2.34 & 4.22 & isolated \\
587725818559266900 & 173.65065 & 67.87359 & 0.06372 & 2.35 & 3.73 & isolated \\
587742014895685727 & 187.70186 & 23.12879 & 0.05658 & 2.34 & 4.14 & central \\
587742189378142355 & 209.30305 & 22.28179 & 0.06239 & 2.46 & 4.14 & isolated \\
587735431226196112 & 191.83299 & 50.81101 & 0.06761 & 2.33 & 4.23 & isolated \\
587742576979542293 & 185.92914 & 20.00920 & 0.06467 & 2.52 & 3.96 & satellite \\
587742644626981283 & 241.78200 & 12.31178 & 0.06314 & 2.64 & 4.12 & satellite \\
588010879837077636 & 189.75342 &  5.43426 & 0.06494 & 2.39 & 4.12 & isolated \\
588013382747029627 & 195.63777 & 51.77336 & 0.05505 & 2.34 & 4.18 & satellite \\
588013382747488287 & 197.34659 & 51.51128 & 0.05591 & 2.32 & 3.79 & central \\
587738570317037616 & 191.82947 & 15.70990 & 0.06844 & 2.39 & 4.08 & central \\
588017702408421556 & 214.37848 &  9.95920 & 0.05780 & 2.36 & 3.97 & isolated \\
588298662505152685 & 201.99579 & 45.75944 & 0.06056 & 2.40 & 4.24 & central \\
588017110221652100 & 182.26642 & 46.81030 & 0.06520 & 2.51 & 4.21 & isolated \\
588017703468269695 & 182.07571 & 12.47228 & 0.06573 & 2.35 & 3.98 & ...     \\
588018089468231733 & 235.36519 & 33.89777 & 0.06992 & 2.34 & 3.94 & isolated \\
588297863639400695 & 129.80997 & 31.34774 & 0.06833 & 2.44 & 3.76 & isolated \\
\cutinhead{{\em NUV-r} red spirals}
587726100412432418 & 219.82480 &  3.36835 & 0.02796 & 2.34 &  5.93 & satellite \\
587725040092250278 & 180.79454 & -2.87600 & 0.05225 & 2.57 &  5.51 & isolated \\
587733603186966558 & 229.87572 & 49.50642 & 0.03706 & 2.42 &  6.94 & central \\
587734622163042661 & 119.56116 & 24.06126 & 0.04469 & 2.31 &  6.21 & isolated \\
587729409146224828 & 237.86009 & 53.16682 & 0.04804 & 2.60 &  7.06 & central \\
587730847425364576 & 315.52719 &  0.20233 & 0.05041 & 2.40 &  6.25 & central \\
587735696987324568 & 212.42142 & 54.90093 & 0.04187 & 2.51 &  5.59 & satellite \\
587736915686130061 & 237.03278 &  9.16713 & 0.03908 & 2.37 &  5.88 & central \\
587732050018238665 & 139.07539 & 47.90812 & 0.05167 & 2.49 &  5.98 & isolated \\
587732470920118561 & 120.32710 & 27.21264 & 0.04702 & 2.47 &  5.91 & central \\
588016892785524828 & 153.39217 & 38.84320 & 0.02123 & 2.63 &  6.44 & satellite \\
587738953105473543 & 206.47536 & 36.56594 & 0.02652 & 2.39 &  5.50 & central \\
587733081884786775 & 176.70801 & 55.70717 & 0.05038 & 2.32 & 13.54 & satellite \\
587739504477733125 & 205.70813 & 29.76290 & 0.04308 & 2.51 &  5.50 & satellite \\
587735347483181134 & 142.45491 & 10.46730 & 0.05103 & 2.54 &  5.50 & isolated \\
587739708476424255 & 197.51146 & 30.04338 & 0.03522 & 2.52 &  5.94 & satellite \\
587739810488909894 & 218.61824 & 23.30360 & 0.03893 & 2.41 &  5.95 & central \\
587739814778110198 & 236.26030 & 21.56277 & 0.04175 & 2.35 &  6.03 & isolated \\
587736899037757746 & 242.20284 & 28.46129 & 0.05022 & 2.48 &  5.74 & satellite \\
587729160052998328 & 224.11003 &  4.53832 & 0.06355 & 2.55 &  5.78 & isolated \\
587729970178097500 & 221.38170 & -2.72378 & 0.05755 & 2.46 &  6.17 & satellite \\
587731870702174485 & 158.12863 & 53.07374 & 0.06347 & 2.60 &  6.68 & satellite \\
587739648355663927 & 167.08467 & 32.00581 & 0.04839 & 2.39 &  5.68 & satellite \\
588011219672236116 & 205.87677 & 61.83131 & 0.03294 & 2.72 &  6.34 & central \\
587732484360372371 & 184.98714 & 50.81934 & 0.06276 & 2.35 &  5.98 & isolated \\
587741489301487648 & 177.11200 & 29.10807 & 0.05186 & 2.62 &  5.85 & central \\
587732771573334191 & 138.50475 &  7.05050 & 0.05627 & 2.45 &  5.92 & satellite \\
587725775606579433 & 118.78078 & 45.72655 & 0.05084 & 2.66 &  5.76 & satellite \\
588017978896482485 & 195.29579 & 39.74447 & 0.03700 & 2.36 &  5.51 & satellite \\
587726032263577792 & 214.35213 &  1.96213 & 0.05257 & 2.46 &  5.80 & satellite \\
588017991237369969 & 229.13567 &  7.03831 & 0.03948 & 2.44 &  5.78 & satellite \\
587735348016775499 & 134.92911 &  9.38926 & 0.06344 & 2.59 &  6.92 & isolated \\
587742015439437970 & 204.78951 & 22.44459 & 0.05309 & 2.54 &  6.14 & central \\
587736541481009403 & 214.25798 &  8.02241 & 0.05939 & 2.50 &  5.34 & satellite \\
587736752468197721 & 241.10870 & 33.18145 & 0.06033 & 2.69 &  7.33 & satellite \\
587736899038543916 & 243.86180 & 27.56479 & 0.06418 & 2.53 &  5.54 & satellite \\
587728917913927863 & 227.05147 & 57.02774 & 0.06764 & 2.69 &  6.28 & isolated \\
587731869092085971 & 160.63490 & 52.28495 & 0.06585 & 2.62 &  5.73 & isolated \\
587731869095231619 & 172.22585 & 54.11234 & 0.06805 & 2.34 &  5.32 & satellite \\
588017625087737918 & 203.95618 & 42.16649 & 0.05378 & 2.50 &  5.39 & isolated \\
587739609171886232 & 205.24374 & 30.89621 & 0.06251 & 2.41 &  5.36 & isolated \\
588017712050471050 & 214.81900 & 48.06102 & 0.05223 & 2.44 & 10.07 & isolated \\
588017724947890388 & 209.22626 &  6.49559 & 0.05447 & 2.51 &  5.35 & isolated \\
587732483300917353 & 229.79492 & 40.04821 & 0.06516 & 2.78 &  6.88 & isolated \\
587732577235828927 & 159.52048 &  5.96323 & 0.06823 & 2.69 &  6.26 & isolated \\
587739811028271306 & 224.75024 & 22.10252 & 0.06195 & 2.42 &  6.21 & isolated \\
588017991236911371 & 228.09845 &  7.30786 & 0.04657 & 2.63 &  5.98 & satellite \\
588018055665746465 & 252.35005 & 26.58406 & 0.05451 & 2.39 &  5.93 & central \\
587733080811896893 & 180.22723 & 55.08683 & 0.06568 & 2.37 &  6.37 & isolated \\
587733397027094667 & 223.01466 & 47.07780 & 0.06952 & 2.62 &  5.58 & isolated \\
587725471207260441 & 131.36423 & 53.25832 & 0.06168 & 2.68 &  5.90 & isolated \\
587725818574799083 & 243.84148 & 52.55205 & 0.06308 & 2.93 &  5.78 & satellite \\
587742013274259567 & 161.30330 & 20.18430 & 0.05525 & 2.51 &  5.50 & satellite \\
587735348030472386 & 166.53897 & 14.01830 & 0.06625 & 2.44 &  8.18 & satellite \\
587742627984834636 & 215.00310 & 15.05922 & 0.06013 & 2.50 &  5.37 & isolated \\
587742889435922717 & 244.29845 & 49.96442 & 0.05744 & 2.44 &  5.61 & central \\
588007005770088667 & 120.90433 & 44.25671 & 0.06339 & 2.68 &  8.22 & isolated \\
588007005802463579 & 243.59914 & 49.21794 & 0.06123 & 2.51 &  5.86 & satellite \\
587736620399001888 & 241.88342 & 27.52636 & 0.06515 & 2.58 &  7.02 & satellite \\
587736620400967937 & 245.86464 & 24.87571 & 0.06775 & 2.60 &  6.09 & satellite \\
588011123574898770 & 175.78755 & 62.08648 & 0.06277 & 2.54 &  6.39 & central \\
588015508211564699 &  36.48389 & -0.81192 & 0.05871 & 2.59 &  6.49 & ...     \\
588017116122644590 & 199.62503 & 47.19300 & 0.05817 & 2.47 &  5.95 & satellite \\
588017625076596923 & 169.22841 & 42.74136 & 0.06492 & 2.71 &  6.93 & satellite \\
588017627770585209 & 199.94577 & 45.00518 & 0.06192 & 2.70 &  5.43 & isolated \\
587739380997685425 & 241.58032 & 21.07602 & 0.06996 & 2.31 &  7.56 & isolated \\
588018055649165425 & 212.49449 & 49.05573 & 0.06368 & 2.54 &  5.57 & isolated \\
588023721781821471 & 220.13477 & 21.44180 & 0.06208 & 2.52 &  6.24 & ...     \\
588297864726052991 & 163.11249 & 45.58035 & 0.06410 & 2.65 & 11.16 & satellite \\
587739651571253552 & 239.04921 & 20.12211 & 0.06909 & 2.53 &  6.73 & isolated \\
587739828204142802 & 221.09988 & 22.98844 & 0.06644 & 2.50 &  6.10 & isolated \\
587739845917868232 & 212.06062 & 23.38638 & 0.06898 & 2.49 &  6.18 & satellite \\
587741708326273396 & 127.92294 & 15.32915 & 0.06603 & 2.45 &  5.79 & isolated \\
587726033325129834 & 186.48831 &  3.00561 & 0.06966 & 2.51 &  7.05 & isolated \\
587742013819519101 & 181.69269 & 22.31201 & 0.06675 & 2.42 &  8.75 & satellite \\
587742014352523427 & 172.27428 & 22.32431 & 0.06936 & 2.57 &  6.71 & isolated \\
587742627458449767 & 239.01050 &  9.72624 & 0.06829 & 2.75 &  6.83 & satellite \\
588007005800562938 & 239.06937 & 52.56771 & 0.06768 & 2.48 &  5.73 & central \\
588009370154696809 & 186.24730 & 61.40760 & 0.06821 & 3.09 &  7.78 & satellite \\
588010360151081078 & 159.53583 &  5.34199 & 0.06722 & 2.41 &  6.26 & satellite \\
588013384356003952 & 189.75845 & 53.30302 & 0.06650 & 2.61 &  7.98 & satellite \\
588017704016609467 & 208.92004 & 11.74395 & 0.06504 & 2.55 &  6.05 & central \\
588017729228046455 & 191.48267 &  7.34779 & 0.06873 & 2.43 &  6.09 & satellite \\
588017947745714223 & 198.05276 & 41.19839 & 0.06702 & 2.42 &  5.38 & isolated \\
588017992297807947 & 198.46162 & 10.40255 & 0.06745 & 2.59 &  5.90 & isolated \\
588023669172469984 & 185.77351 & 19.82585 & 0.06773 & 2.33 &  5.31 & central \\
\enddata
\tablecomments{Column(1): photometric identification number from SDSS DR7.
Column (2): Right Ascension from SDSS.
Column (3): Declination from SDSS.
Column (4): redshift from SDSS.
Columns (5)-(6): {\em u-r} and {\em NUV-r} color measured within the
{\em r}-band $\mu_r = 25\, {\rm mag\,arcsec^{-2}}$ apertures,
corrected for the foreground Galactic extinction.
Column (7): environment classification from \citet{Yang2007}.}
\label{table1}
\end{deluxetable}

\subsection{Parameter Derivation\label{sec:parameter}}

For the purpose of understanding the differences between {\em NUV-r} blue and
red spirals, we will visit the star formation main sequence relation, the
spatial distribution of the {\em NUV} and optical emission, the bulge and
disk mass-size relations and the environment in this paper.
Except for the surface brightness profiles in the {\em NUV} and optical
wavebands, the parameters were primarily drawn from the public data release
and are similar to those used in \citet{Guo2020}, for which we only give brief
descriptions of how they were derived below.

The bulge masses, disk masses and total stellar masses were extracted from
\citet{Mendel2014}. Briefly, two-dimensional image decompositions were
performed on SDSS $u$-, $g$-, $r$-, $i$- and $z$-band images \citep{Simard2011,
Mendel2014}, under the assumption of both a de Vaucouleurs bulge + disk model
and a single-component S{\'e}rsic model. Based on the best-fit models, spectral
energy distributions (SEDs) across the five wavebands were fitted to derive the
best estimates of stellar masses.  The {\em r-}band sizes for the bulge and
disk components were also derived from the two-dimensional image decompositions
\citep{Simard2011}.  We adopted the stellar masses and sizes of bulges and
disks exclusively for those bulge + disk systems, which consist of 97.9\%
(46/47) {\em NUV-r} blue spirals and 83.7\% (72/86) {\em NUV-r} red spirals.

We derived the {\em NUV} band and optical {\em u-}, {\em g-} and {\em r-} bands
surface brightness profiles (SBPs) based on the {\it GALEX} and SDSS images,
drawn from the NSA. These images have been background subtracted.  The
remaining processes used to extract the SBPs include the generation of mask
images and the surface photometry. The mask images were obtained by running
SExtractor \citep{Bertin1996}, and the surface photometry was carried out using
the IRAF/ellipse task. We used series of fixed-shape concentric elliptical
annuli with a step of 1 pixel in radius to extract the SBPs. The ellipticities
and azimuth angles of the elliptical annuli were obtained from the NSA
catalogue. We found that the {\em NUV} SBPs can extend to three times the
effective radius ($R_{\rm e}$) with good quality, where $R_{\rm e}$ is
represented by the {\em r-}band elliptical Petrosian 50\% light radius from the
NSA.

The PSFs vary across different wavebands. For the SBPs in different wavebands,
we retained the original spatial resolution of the images. When deriving the
color profiles, the image with the narrower PSF was convolved with a Gaussian
kernel to match the broader PSF in the other waveband before performing
photometry. We adopted a PSF size of 5.3\arcsec\ for the {\it GALEX} {\em NUV}
band \citep{Morrissey2007}, and obtained the PSF widths of the SDSS wavebands
from the tsField files. Furthermore, for a fair comparison with the {\em
NUV} emission, the optical SBPs and the color profiles were truncated beyond 3
$R_{\rm e}$.

The SFRs were adopted from the {\it GALEX}-SDSS-{\it WISE} Legacy Catalog
All-sky 2 \citep[GSWLC-A2;][]{Salim2018}, which includes the SFRs obtained
using Bayesian SED fitting based on the UV, optical and MIR observations from
{\it GALEX}, SDSS and {\it WISE}, respectively. There are 83.0\% (39/47) of
{\em NUV-r} blue spirals and 86.0\% (74/86) of {\em NUV-r} red spirals with SFR
measurements. The SFR error increases with decreasing SFR.
For {\em NUV-r} blue spirals, the errors are mostly
below 0.2 dex, except for several galaxies with $\log \SFR < -0.6 $, whose SFR errors
can be as large as 0.5-0.8 dex. For {\em NUV-r} red spirals, the SFR errors range
from 0.2 to 1 dex.

Group memberships and central-satellite classifications were taken from
\citet{Yang2007}. There are 97.9\% (46/47) of {\em NUV-r} blue spirals and
97.7\% (84/86) of {\em NUV-r} red spirals included in the group catalog. In
this catalog, galaxies were either classified as a central galaxy or identified
as a satellite. As a result, isolated galaxies were not distinguished from
truly central galaxies located in galaxy groups or clusters. To better
understand the surrounding environments of our sample galaxies, we
differentiated central galaxies without satellites, i.e., isolated galaxies,
from centrals with at least one satellite galaxy.
The environment classifications for the {\em NUV-r} blue and red spirals
are listed in Table \ref{table1}.

\section{RESULTS\label{sec:results}}

In this work, we focus on investigating the differences between optically red,
massive spiral galaxies with different {\em NUV-r} colors and their underlying
driving mechanisms. For this purpose, we will compare the star formation
stages, the light distributions in the {\em NUV} and optical wavebands, the
structural properties, as well as the environments of our samples of {\em
NUV-r} blue and red spirals.

\subsection{Global Star Formation Properties \label{subsec:MS}}

\begin{figure}
\plotone{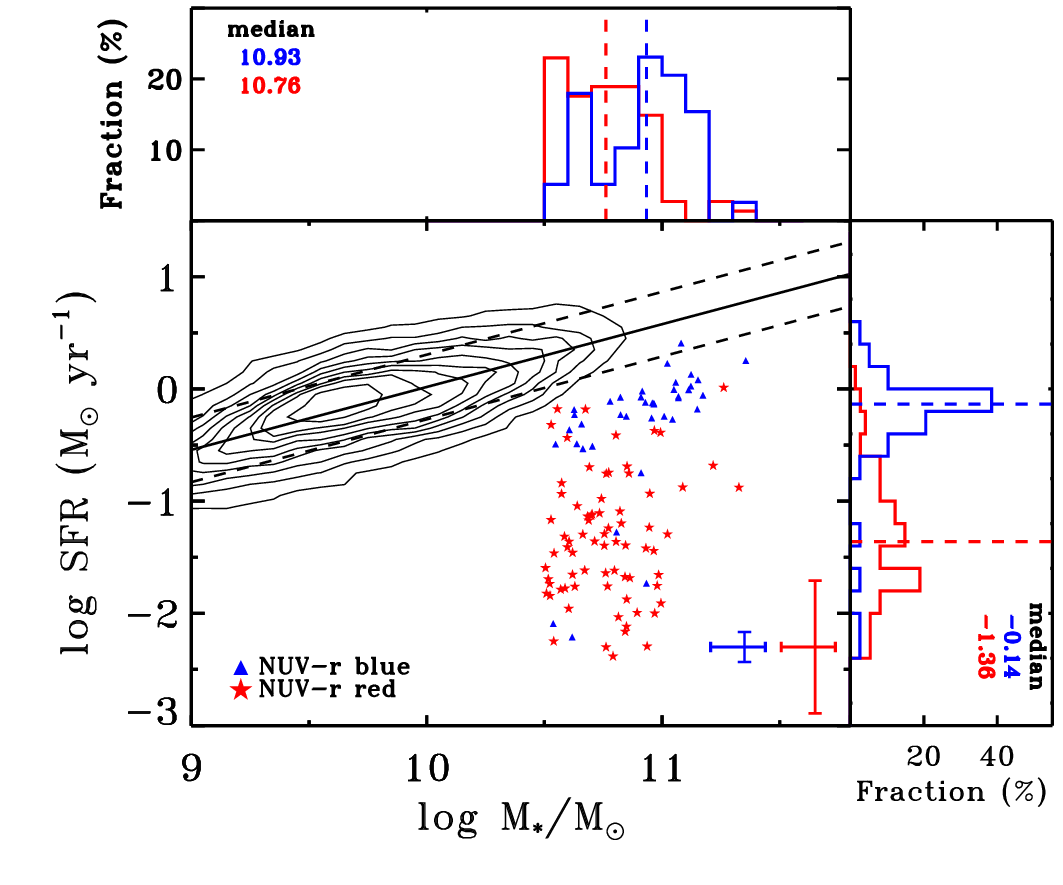}
\caption{SFR vs. stellar mass relation for optically red spirals with
blue (blue triangles) and red (red stars) {\em NUV-r} colors.
The contours show the number density distribution of a parent sample of
galaxies with $0.02 < z < 0.07$ in the catalog of \citet{Mendel2014}.
The black solid line represents the main sequence
(${\rm log\, SFR} = 0.56 \times {\rm log} (M_*/M_{\odot})\, -\, 5.58$)
with $\sim$ 0.3 dex scatter
(dashed lines) defined by the parent sample.
The error bars in the bottom-right corner are the median measurement errors
for {\em NUV-r} blue and red spirals, respectively.
The top- and right-hand panels show the histograms of the stellar mass and
SFR for each sample with dashed lines representing the corresponding
median values.
{\em NUV-r} red spirals are fully quenched, with SFRs 1-3 dex lower than
those of MS galaxies, while {\em NUV-r} blue spirals still exhibit weak star
formation, with a median sSFR of $8.09 \times 10^{-12} {\mathrm{ yr^{-1}}}$.
}
\label{ms.ps}
\end{figure}

The correlation between stellar mass and SFR for star-forming galaxies, known
as the star formation main sequence (SFMS) relation, has led to the widespread
use of the SFR versus stellar mass diagram to evaluate the star formation
status of galaxies. We explore the positions of {\em NUV-r} blue and red spiral
galaxies on the SFR vs. stellar mass diagram in Figure \ref{ms.ps}. The
contours in Figure \ref{ms.ps} show the number density distribution of
star-forming galaxies in the parent sample with $0.02 < z < 0.07$ from
\citet{Mendel2014}. The black solid line represents the best-fit SFMS relation
based on the parent sample, with a scatter of $\sim$ 0.3 dex, denoted by the
dashed lines. It is obvious that {\em NUV-r} blue and red spirals populate
different regions in the SFR vs. stellar mass diagram. The vast majority
of {\em NUV-r} red spirals are located far below the SFMS relation, with SFRs
1-3 dex smaller than the MS galaxies, suggesting that {\em NUV-r} red spirals
have been fully quenched. In contrast, the {\em NUV-r} blue spirals are 
mostly $\sim$ 0.5-0.9 dex lower than the SFMS ridgeline, with a median
sSFR of $8.09 \times 10^{-12} {\mathrm{ yr^{-1}}}$, indicating that the
optically red but {\em NUV-r} blue spirals are still processing weak star
formation. Such a difference in SFR is further presented in the histogram shown
in the right-hand panel of Figure \ref{ms.ps}. Hence, the {\em NUV-r} blue
spirals do not belong to the population of truly passive galaxies. These
results are consistent with the argument that optically selected red spirals
can be contaminated by galaxies with low levels of star formation, as discussed
in Section \ref{sec:intro}.

\subsection{Radial distribution of {\em NUV} emission \label{subsec:SBP}}

The {\em NUV-r} blue and red spirals are classified according to their global
{\em NUV-r} colors. In this subsection, we will compare the spatial and radial
distributions of the NUV and optical emission to identify the regions where
the differences appear.

\begin{figure}
\plotone{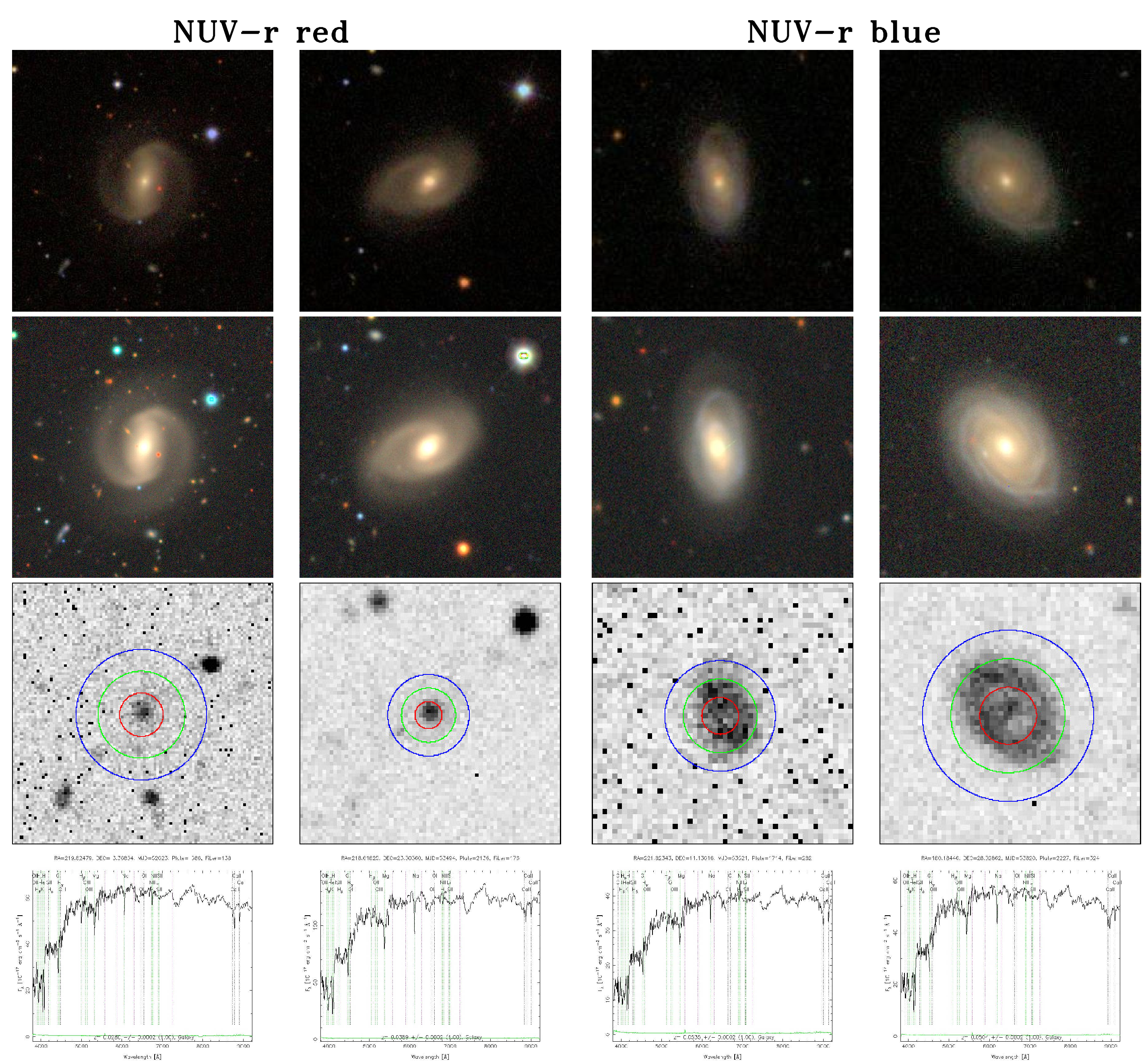}
\caption{
True color SDSS images (first row), DESI images (second row), {\it GALEX}
{\em NUV}-band images (third row) and SDSS fiber spectra (fourth row) for
example {\em NUV-r} red (left two columns) and blue (right two columns)
spirals, respectively.
The physical size of each image is 80 $\times$ 80 kpc$^2$.
The red, green and blue circles on the {\it GALEX} images represent 1, 2, and 3
$R_{\rm e}$.
{\em NUV-r} blue spirals exhibit intense {\em NUV} emission in their outer regions.
}
\label{imgSDSSGALEX.ps}
\end{figure}

In Figure~\ref{imgSDSSGALEX.ps}, we show the true color images from SDSS and
the DESI Legacy Imaging Survey \citep{Dey2019}, as well as the {\it GALEX} {\em
NUV} images for four example galaxies in the {\em NUV-r} red and blue spiral
samples, respectively. It is obvious that the deeper DESI images are able to
capture more diffuse emission in the outskirts of galaxies, and show
clearer structures than the SDSS images. For {\em NUV-r} blue spirals, the blue
structures observed in the DESI images exhibit a stark contrast to their red
inner regions, whereas the corresponding regions in the SDSS images do not show
such a significant difference from the inner parts of galaxies. By comparing
the DESI optical images of these two types of spirals in Figure
\ref{imgSDSSGALEX.ps}, we can see that {\em NUV-r} blue spirals have bulge
colors similar to those of {\em NUV-r} red spirals, but their disk colors
differ distinctly. Such optical color distributions are in well agreement with
the NUV emission features, as shown in the third row of Figure
\ref{imgSDSSGALEX.ps}.  The {\em NUV-r} red spirals exhibit almost no {\em NUV}
emission, except for the faint {\em NUV} emission in their central regions
(within 1\,$R_{\rm e}$), which likely originates from old stellar populations
rather than star formation, as evidenced by the large $4000\AA$ breaks and
the absence of H$\alpha$ emission lines in the optical fiber spectra exhibited
in the bottom row of Figure~\ref{imgSDSSGALEX.ps}. In comparison, {\em NUV-r}
blue spirals exhibit strong {\em NUV} emission, especially in their outer parts
(between 1-3\,$R_{\rm e}$). 

\begin{figure}
\includegraphics[width=0.48\textwidth]{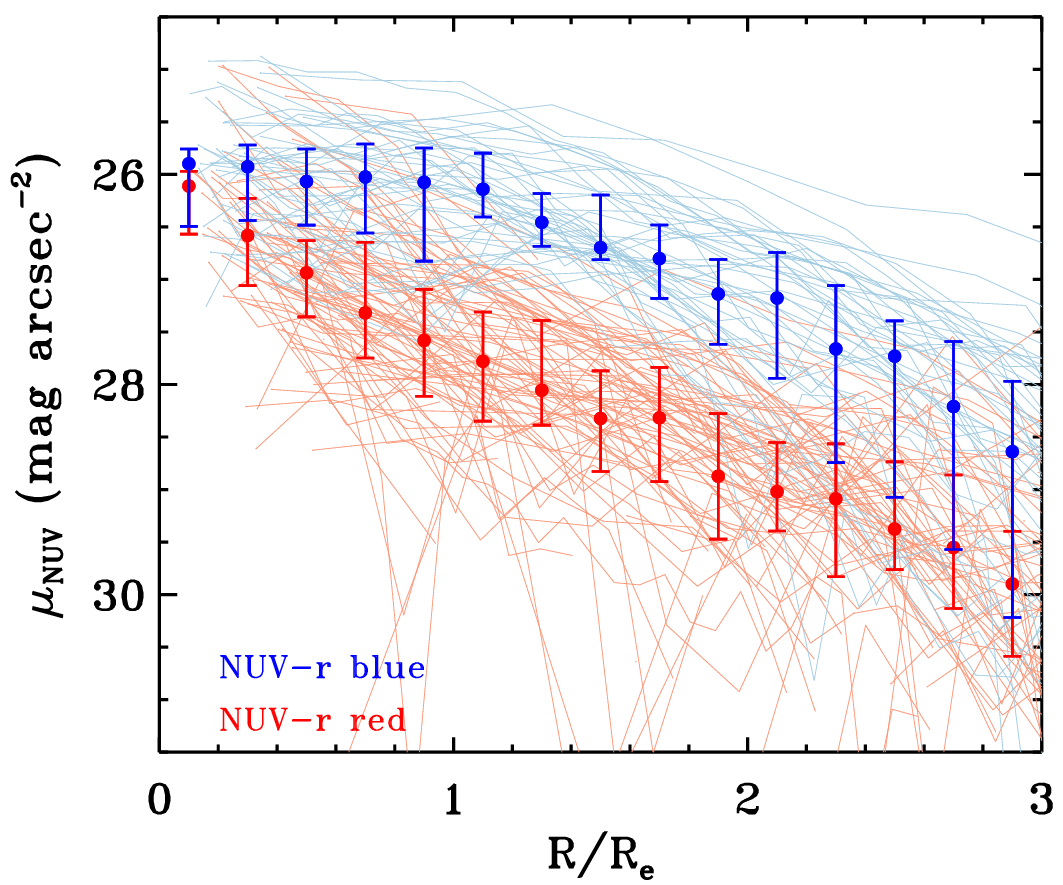}
\includegraphics[width=0.48\textwidth]{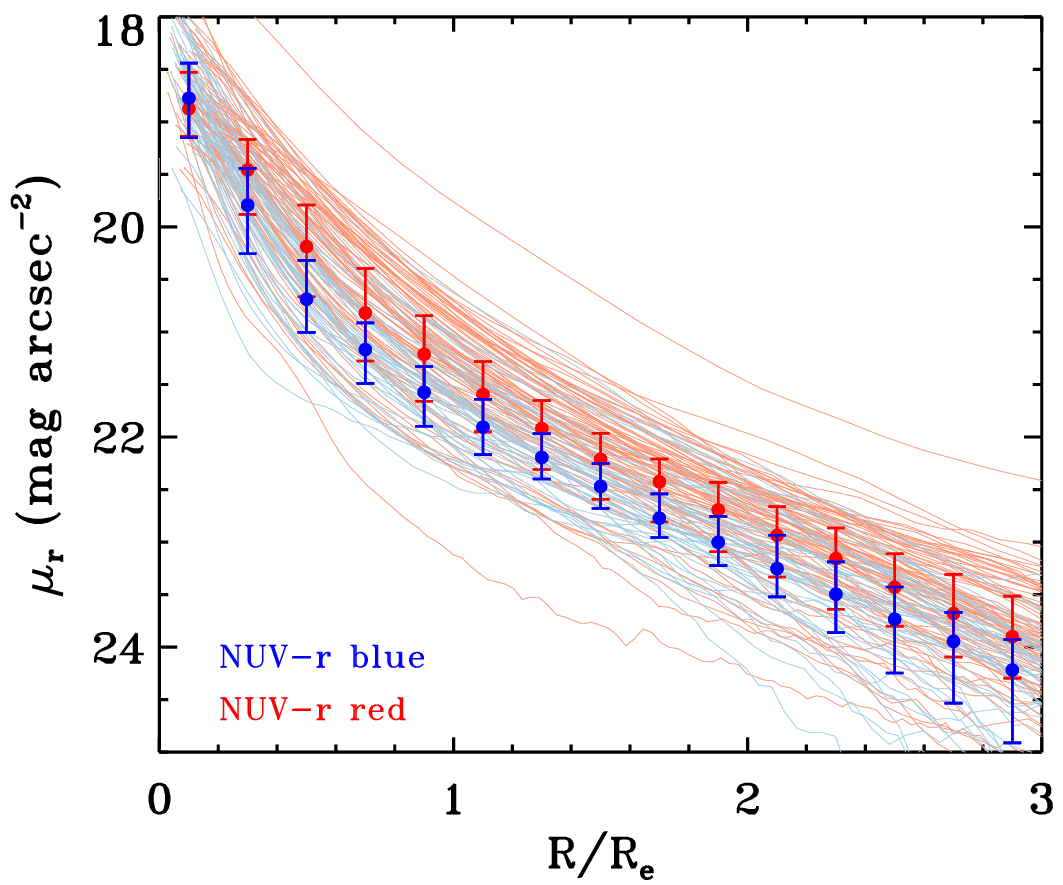}
\caption{{\it GALEX} {\em NUV}-band (left panel) and SDSS {\em r}-band (right panel) SBPs
for {\em NUV-r} blue (blue lines) and red (red lines)
galaxies. Blue and red solid circles are
the median SBPs with errors bars denoting the lower (25\%)
and upper (75\%) quartiles for each sample.
Blue spirals have brighter SBPs
than the red ones in the {\em NUV} band, especially outside 1 $R_{\rm e}$.
While the {\em r}-band SBPs of the {\em NUV-r} blue spirals are fainter than the
{\em NUV-r} red ones, but the difference is comparable to the rms scatters.}
\label{SBP.ps}
\end{figure}

We quantify the {\it GALEX} {\em NUV} and SDSS $r$-band spatial distributions
of {\em NUV-r} blue and red spirals by investigating their SBPs in Figure~\ref
{SBP.ps}. To make a fair comparison among galaxies with different sizes, we
plot the surface brightness as a function of radius normalized to effective
radius ($R/R_{\rm e}$). The left panel of Figure~\ref {SBP.ps} displays the
{\em NUV}-band SBPs for all sample galaxies.  The solid
circles with error bars represent the median SBPs, and the thin curves denote
individual profiles.  It is clear that both the median and individual SBPs of
{\em NUV-r} blue spirals are significantly different from those of {\em NUV-r}
red spirals. The median surface brightness of {\em NUV-r} red galaxies
decreases monotonically from the center to $\sim$ 3\,$R_{\rm e}$, at which the
surface brightness approaches $\sim$ 30\,$\magsec$. By comparison, the median
surface brightness of {\em NUV-r} blue galaxies shows a roughly flat profile
within $\sim 1\, {\rm R}_{e}$ before decreasing to $\sim$ 28.5\,$\magsec$
by $\sim$ 3\,$R_{\rm e}$.  Overall, the median surface brightness of {\em
NUV-r} blue galaxies is $\sim$ 1.5-2\,$\magsec$ brighter than that of
{\em NUV-r} red galaxies from 1\,$R_{\rm e}$ to 3\,$R_{\rm e}$, indicating that
star formation mainly takes place in the outer disks (1-3\,$R_{\rm e}$) of {\em
NUV-r} blue spirals, consistent with the examples shown in
Figure~\ref{imgSDSSGALEX.ps}.  In contrast, the $r$-band SBPs shown in the
right panel of Figure~\ref {SBP.ps} present that the {\em NUV-r} blue spirals are
fainter than their {\em NUV-r} red counterparts, but the difference
is comparable to the rms scatters. 

\begin{figure}
\plotone{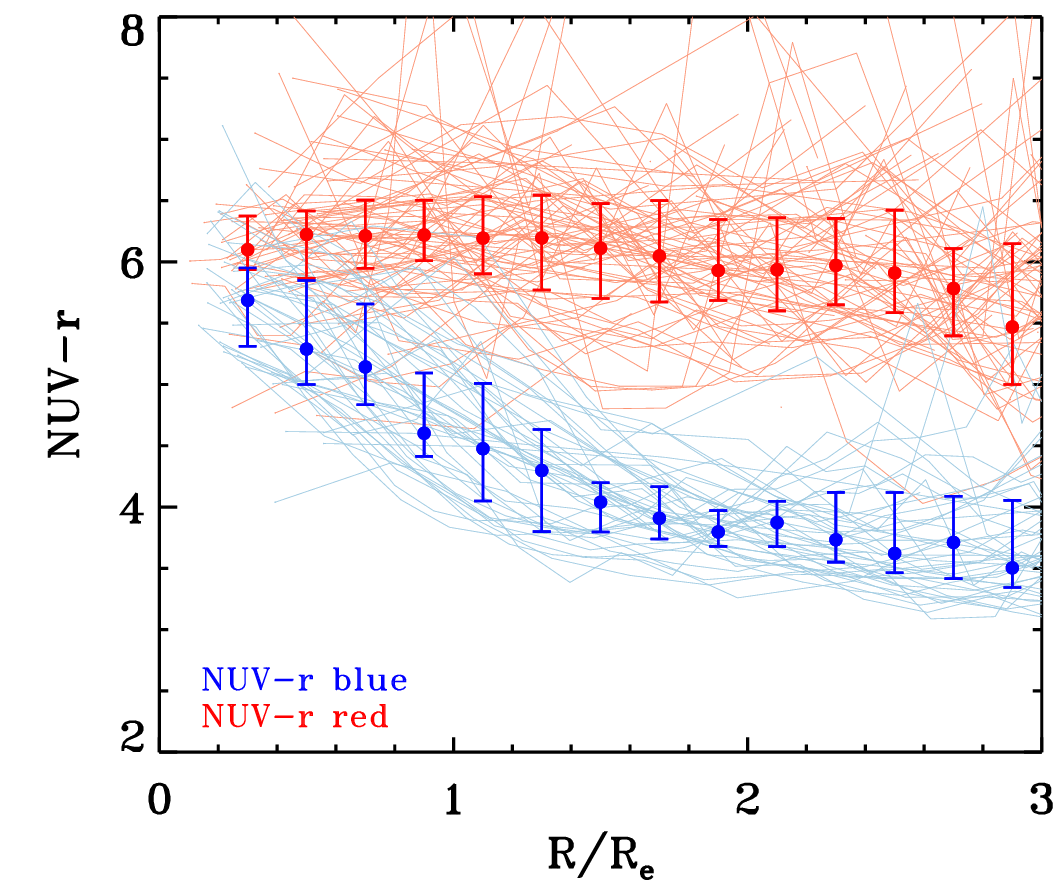}
\caption{{\em NUV-r} color profiles for {\em NUV-r} blue (blue lines) and red
(red lines) galaxies.
Blue and red solid circles are the median color profiles with errors bars
denoting the lower (25\%)
and upper (75\%) quartiles for each sample.
{\em NUV-r} red spirals exhibit a consistent red color from
the centers to the outskirts, whereas {\em NUV-r} blue spirals are red within
$\sim$ 0.8\,$R_{\rm e}$, yet transition to the blue region ($NUV-r < 4$) beyond
$\sim$ 1.5 $R_{\rm e}$.}
\label{colorPF.ps}
\end{figure}

A direct contrast between the NUV and $r$-band light distributions is the {\em
NUV-r} color profile. Figure \ref{colorPF.ps} compares the {\em NUV-r} color
profiles for our {\em NUV-r} blue and red sample galaxies. It is obvious
that the {\em NUV-r} red spiral galaxies have $NUV-r > 5$ from the center to
$\sim$ 3\,$R_{\rm e}$, suggesting that they have been quenched completely.
However, {\em NUV-r} blue spirals have $NUV-r > 5$ only within $\sim$
0.8\,$R_{\rm e}$, which is approximately 2.1 times the bulge $R_{\rm e}$
on average, then become bluer and fall into the region for {\em NUV-r} blue
spirals ($NUV-r < 4$) outside $\sim$ 1.5\,$R_{\rm e}$. This indicates
that {\em NUV-r} blue spirals host quenched bulges and inner disks, as well as
star-forming outer disks. The region in between (i.e., 0.8-1.5\,$R_{\rm
e}$), with $4 < NUV-r < 5$, represents a transition from red inner
disks to star-forming outer disks. The red and dead bulges dominate the
integrated optical luminosities, which makes the global optical colors of {\em
NUV-r} blue spirals red. On the other hand, the star-forming outer regions lead
to their blue global {\em NUV-r} colors.

\begin{figure}
\includegraphics[width=0.48\textwidth]{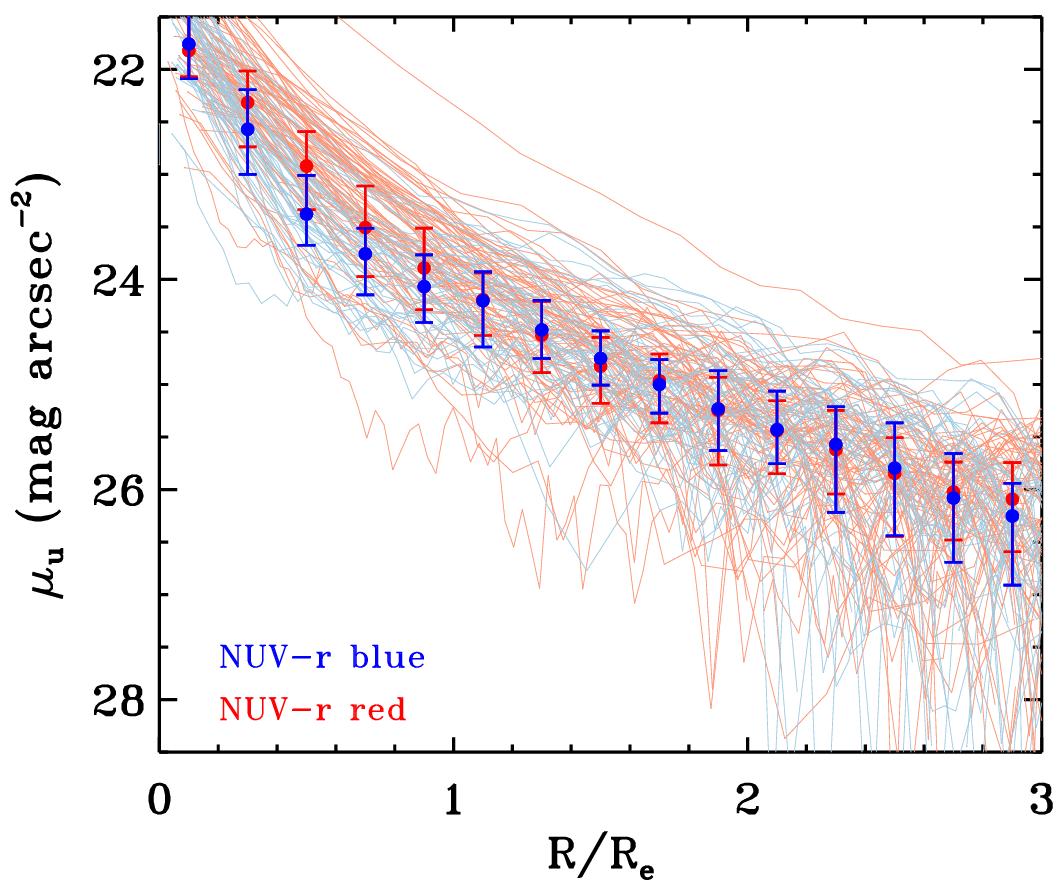}
\includegraphics[width=0.48\textwidth]{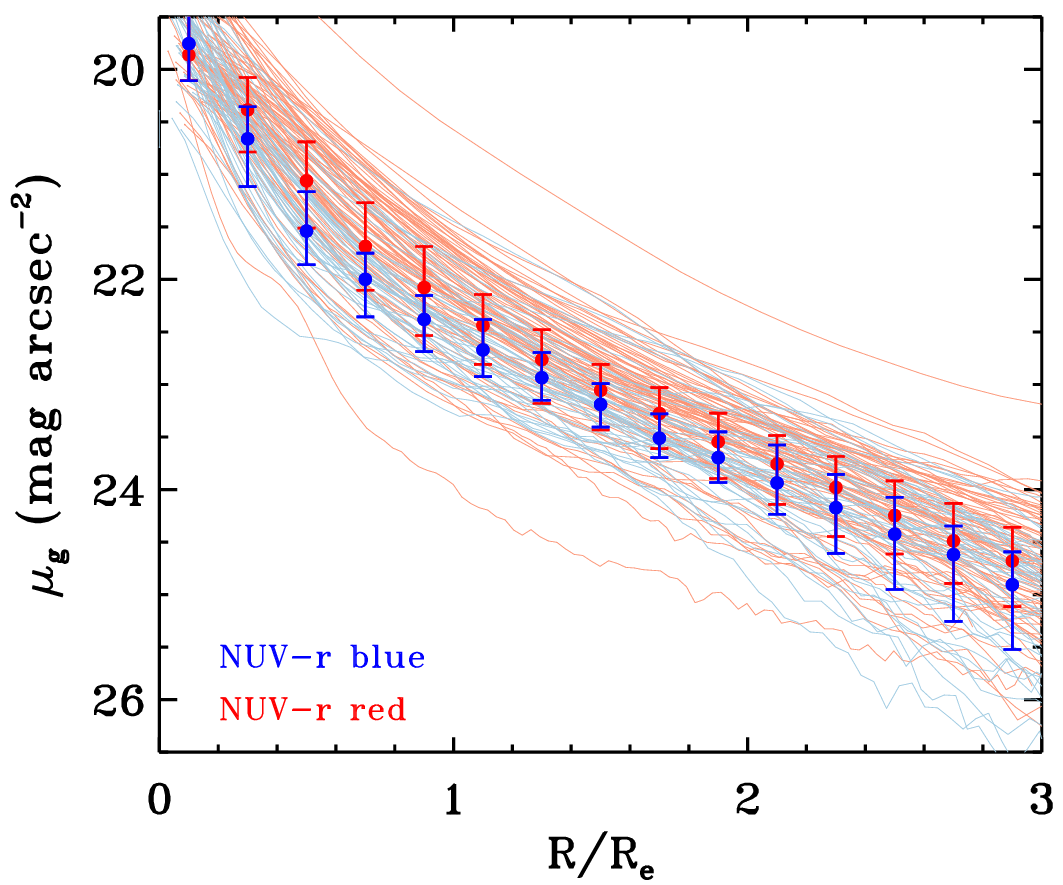}
\\
\includegraphics[width=0.48\textwidth]{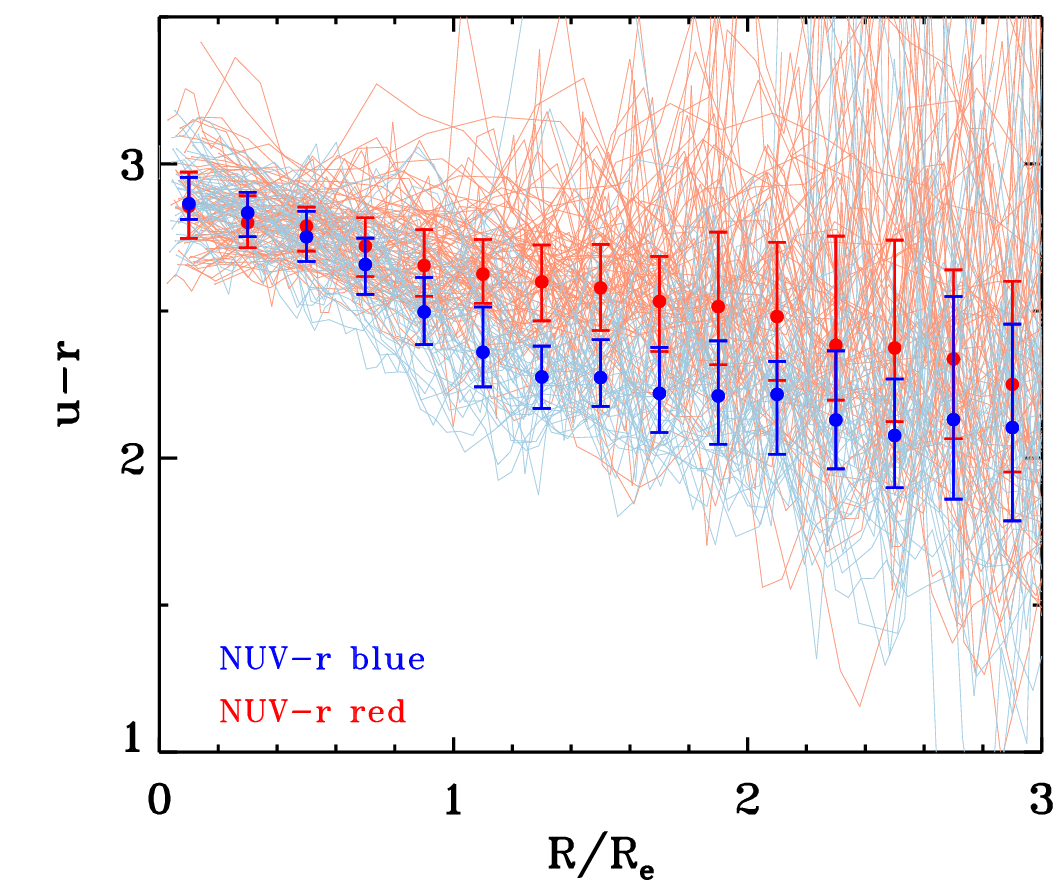}
\includegraphics[width=0.48\textwidth]{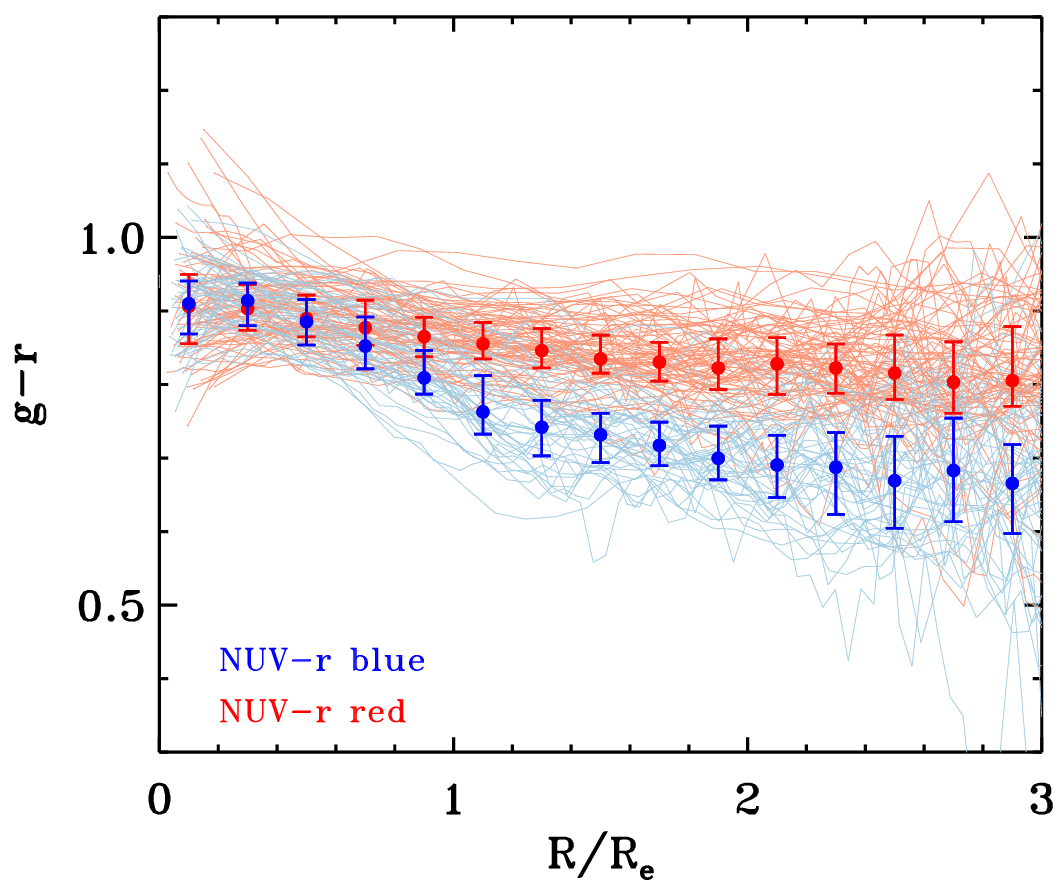}
\caption{
SDSS {\em u-}band (upper-left panel) and {\em r-}band (upper-right panel)
SBPs, along with {\em u-r} (lower-left panel) and {\em g-r} (lower-right panel)
color profiles, for {\em NUV-r} blue and red spirals.
The symbols are the same as in Figure \ref{SBP.ps}.
The color difference between {\em NUV-r} blue and red spirals only becomes
distinguishable beyond 1 $R_e$.
}
\label{ugprof.ps}
\end{figure}

Considering that our samples of galaxies were originally selected by integrated
{\em u-r} colors, and {\em g-r} colors were often used to select red spirals in
the 2010s, it is interesting to look at the {\em u-} and {\em g-}band SBPs, and
the {\em u-r} and {\em g-r} color profiles, which are shown in Figure~\ref
{ugprof.ps}. We note that the outer parts of the {\em u-}band SBPs suffer from
larger uncertainties due to the fainter {\em u-}band emission relative
to the other SDSS wavebands. The upper panels of Figure~\ref{ugprof.ps}
clearly show that for both {\em u-}band and {\em g-}band SBPs, the {\em NUV-r}
blue and red spirals overlap significantly, with the difference even smaller
than that in the {\em r }band. The {\em u-r} and {\em g-r} color profiles shown
in the lower panels of Figure~\ref{ugprof.ps} further indicate that the color
difference in the optical bands between {\em NUV-r} blue and red spirals only
becomes distinguishable beyond $1\, R_{\rm e}$. This suggests that the outer
optical color profiles can differentiate {\em NUV-r} blue from {\em NUV-r} red
spirals. However, {\em NUV} band is definitely a better indicator of recent
star formation than the optical bands. 

\subsection{Structural Properties\label{subsec:properties}}

\begin{figure}
\includegraphics[width=0.48\textwidth]{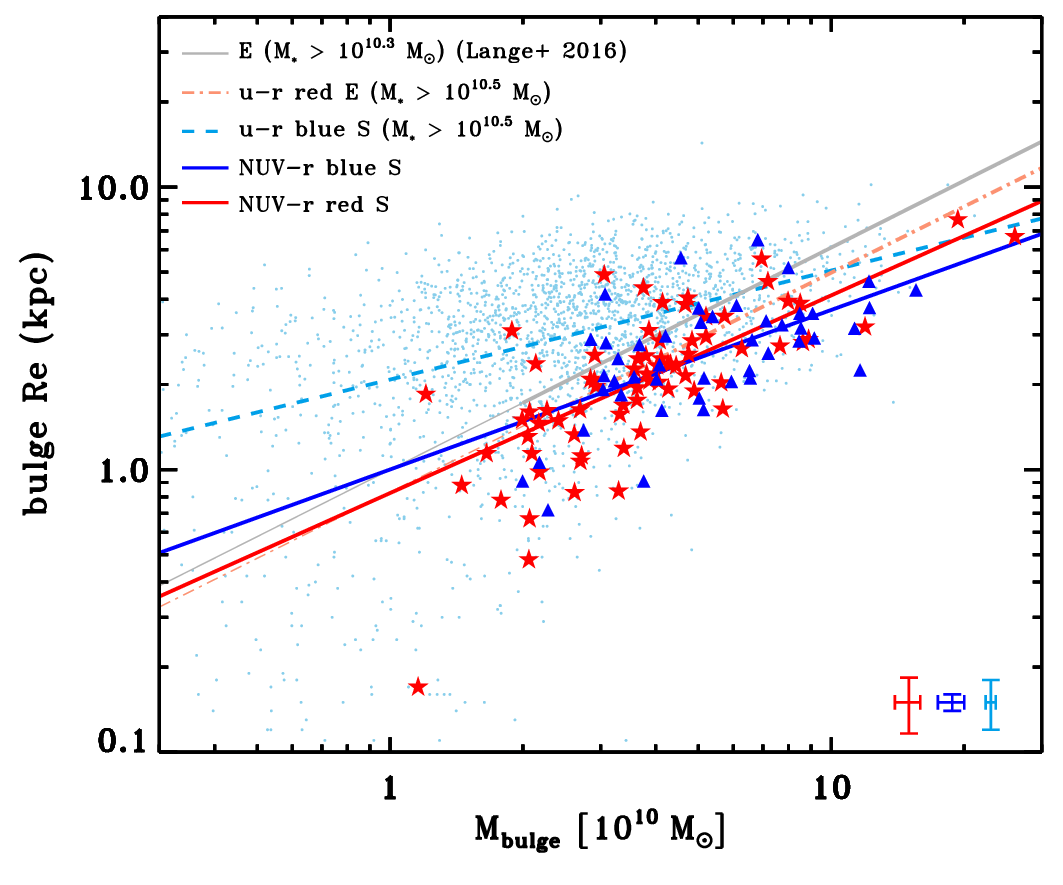}
\includegraphics[width=0.48\textwidth]{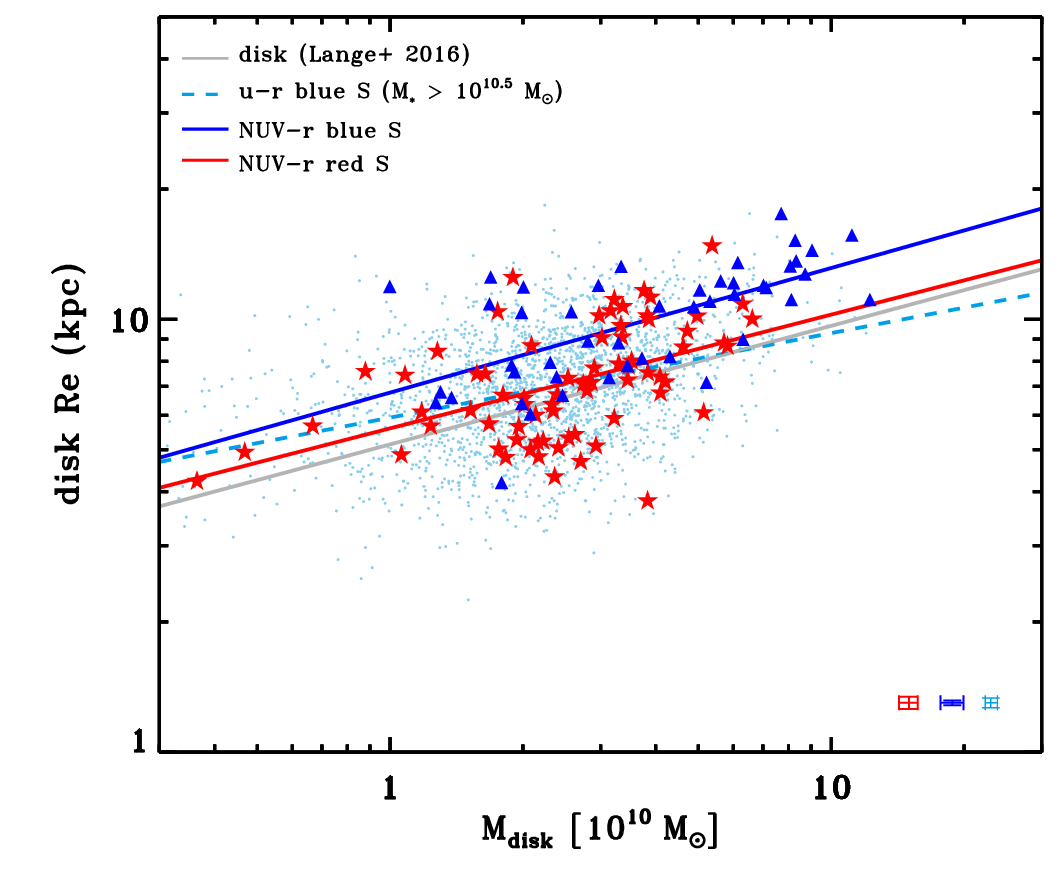}
\caption{
SDSS {\em r}-band size versus stellar mass relation for bulges (left panels) and disks (right
panels) of the {\em NUV-r} blue (blue triangles) and red (red stars) spirals.
The blue and red solid lines show the best-fit relations for the two samples.
The light blue dots are the {\em u-r} blue spirals in our control sample,
and the light blue dashed line illustrates the best-fit relation for these
galaxies.
The gray solid line represents the {\em r}-band relation for local massive ellipticals
($ R_e = 0.999 \times (\frac{M_*}{10^{10}M_{\odot}})^{0.786}$) in the left
panel and local disks
($ R_e = 5.141 \times (\frac{M_*}{10^{10}M_{\odot}})^{0.274}$) in the
right panel from \citet{Lange2016}.
The light red dot-dashed line in the left panel represents the
best-fit relation for {\em u-r} red ellipticals.
The thin lines represent the extrapolated fitting relationships.
The error bars shown in the bottom-right corner in each panel represent the
median measurement errors for the {\em NUV-r} blue and red sample galaxies,
as well as {\em u-r} blue control sample galaxies, respectively.
}
\label{Re_Rd.ps}
\end{figure}

To understand the underlying physical reasons for the differences and
similarities in the bulge and disk colors between {\em NUV-r} blue and red
spirals, we compare the mass-size relations for their bulges and disks
separately. It is known that the size of a galaxy is related to its specific
angular momentum, and hence is a probe of its structure growth history
\citep{Romanowsky2012,Lange2016,vanderwel2014, George2025}. Elliptical galaxies
and spiral galaxies have experienced different formation and evolution
histories, which have imprinted on their different mass-size relations
\citep{Shen2003,vanderwel2014}. Similarly, the bulge and disk components of a
massive galaxy are expected to obey different mass-size relations, since they
share similar evolutionary histories to elliptical and disk galaxies,
respectively. 

Previous studies have separately investigated the mass-size
relations of bulges and disks. For instance, utilizing a bulge + disk
two-component decomposition approach in the {\em r} band, \citet{Lange2016}
obtained different mass-size relations for spheroid and disk components,
based on a sample of $\sim$7,500 galaxies from the Galaxy And Mass
Assembly (GAMA) survey with a broad coverage in Hubble type. While the sample,
data and measurement methods in \citet{Lange2016} differ from those adopted
here, a rough comparison may still be instructive. Furthermore, to facilitate a
fairer comparison, we select reference samples of {\em u-r} blue spirals
($ u-r < -0.673+0.227 \log (M_*/M_{\odot}) $) and
{\em u-r} red ellipticals from the same parent sample as the {\em
u-r} red spirals, which ensures consistent stellar mass and redshift ranges
across all subsamples, and identical methods are applied for mass and size
measurements.

Figure~\ref{Re_Rd.ps} compares the mass-size relations in the {\em r} band for
the bulge (left panel) and disk (right panel) components of the two types of
optically red spirals, and the {\em u-r} blue spirals. Following
\citet{Lange2016}, we fit the mass-size relation for each sample with a power
law model. The best-fit parameters are listed in Table \ref{table2}. We overplot the
best-fit relations from \citet{Lange2016} for comparison. Specifically, the
relation established from the massive elliptical galaxies with $M_* \ge 2\times
10^{10}M_\odot$ by \citet{Lange2016} is adopted for comparison with the bulge
components, while for the disk components, we plot their best-fit relation for
disks built on the combined population of Sd-Irr, late-type galaxies and the
disks of early-type galaxies. In the left panel of Figure~\ref{Re_Rd.ps}, we
also overplot the best-fit relation for our red massive ($M_* \ge
10^{10.5}M_\odot$) ellipticals (the light red dot-dashed line) for comparison.
In spite of the differences in samples and methods, the best-fit mass-size
relation based on our red massive ellipticals is very similar to that derived
by \citet{Lange2016} for their massive ellipticals. 

As can be seen from the left panel of Figure~\ref{Re_Rd.ps}, the {\em NUV-r}
blue and red spirals are well mixed in the bulge mass-size diagram.
Furthermore, they roughly follow the trend of the mass-size relations for
massive ellipticals, indicating that their bulges behave similarly to massive
ellipticals. This agrees with the finding of \citet{FraserMcKelvie2018}
that passive spirals exhibit a high de Vaucouleurs flux fraction ($fracDeV >
0.6$). In comparison, the bulges of {\em u-r} blue spirals follow a shallower
mass-size relation than their red counterparts, which means that at a given
bulge mass, the bulges of {\em u-r} blue spirals tend to be larger. This is
consistent with the finding that {\em u-r} blue spirals have lower central
stellar mass surface densities than {\em u-r} red spirals in \citet{Guo2020}.
The different bulge mass-size relations followed by spirals of varying {\em
u-r} colors suggest that they have experienced different formation processes.

As for the disk components, the right panel of Figure~\ref{Re_Rd.ps} shows
clear difference between {\em NUV-r} blue and red spirals. The {\em NUV-r} red
spirals distribute similarly to the {\em u-r} blue spirals, and are roughly
consistent with the disk mass-size relation derived by \citet{Lange2016}. In
contrast, {\em NUV-r} blue spirals lie above the relations. The best-fit
relations indicate that, at a given disk mass, {\em NUV-r} blue spirals possess
larger disks than their {\em NUV-r} red counterparts, by a factor of $\sim
1.20$ in disk size. This suggests that the {\em NUV-r} red spirals have disk
structures similar to those of the bulk disk population, whereas the {\em
NUV-r} blue spirals possess larger disks. 

\begin{figure}
\plotone{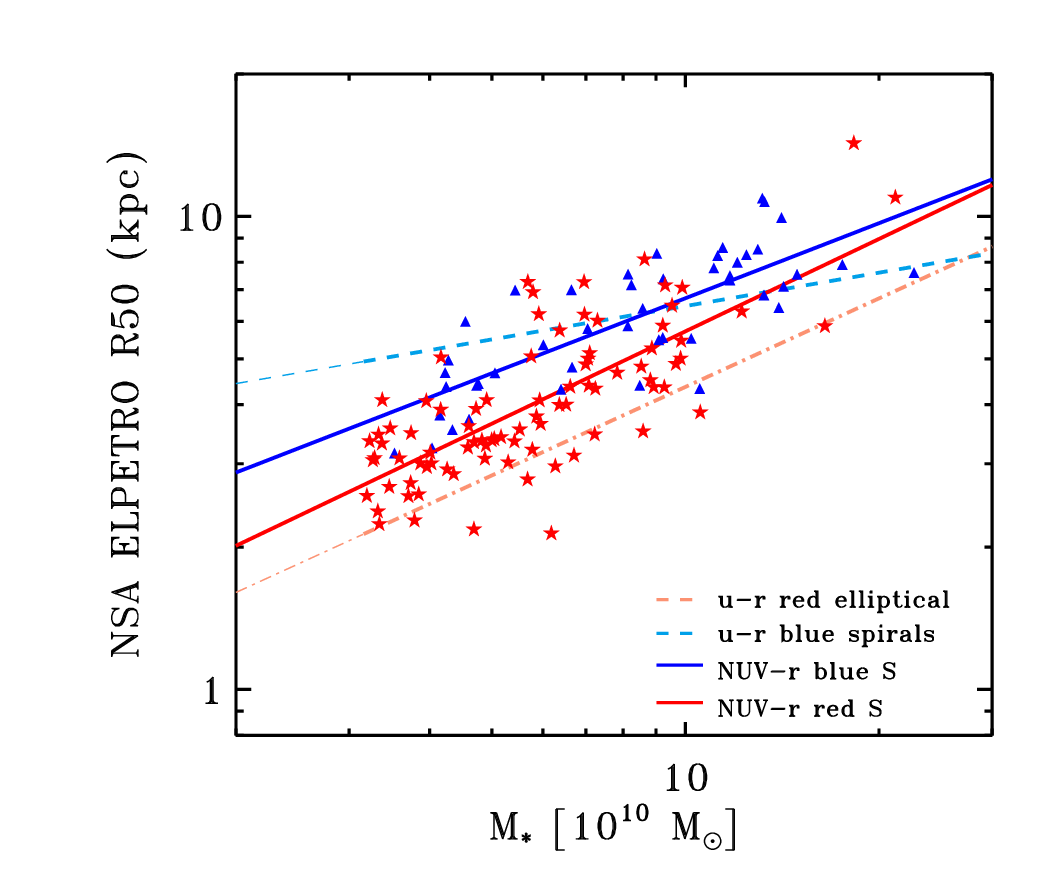}
\caption{
Global stellar mass-size relation for {\em NUV-r} blue (blue triangles) and
red (red stars) spirals.
The blue and red solid lines represent the best-fit relations for the two samples.
The light red dot-dashed line and light blue dashed line represent our control
{\em u-r} red elliptical and blue spiral samples, respectively.
The thin lines show the extrapolated fits.
}
\label{globalmass-size.eps}
\end{figure}

\startlongtable
\begin{deluxetable}{lcc}
\tabletypesize{\footnotesize}
\tablecolumns{3}
\tablewidth{0pt}
\tablecaption{Best-fit parameters to the mass-size relation $R_e = a \times (\frac{M_*}{10^{10}M_{\odot}})^{b}$ for the bulge and disk components}
\tablehead{
\colhead{Case} & \colhead{a} & \colhead{b}}
\startdata
\cutinhead{{\em NUV-r} blue spirals}
bulge & $1.001 \pm 0.208$ & $0.565 \pm 0.109$ \\
disk & $6.773 \pm 0.690$ & $0.288 \pm 0.061$ \\
\cutinhead{{\em NUV-r} red spirals}
bulge & $0.826 \pm 0.107$ & $0.700 \pm 0.079$ \\
disk & $5.600 \pm 0.254$ & $0.263 \pm 0.042$ \\
\enddata
\label{table2}
\end{deluxetable}

Thanks to the multi-component decomposition technique, which enables us to
study the bulge and disk components separately. However, such decompositions
may still experience challenges in cleanly separating these two components. To
test the reliability of the larger disk sizes of {\em NUV-r} blue spirals
relative to their red counterparts, a simple and straightforward way is to
examine the global stellar mass-size relation. As shown in
Figure~\ref{globalmass-size.eps}, {\em NUV-r} blue spirals present larger
overall sizes than their red counterparts at fixed stellar masses, confirming
the trend seen in the disk size comparison.

\subsection{Environments and HI Contents \label{subsec:environment}}

As presented in the previous subsections, {\em NUV-r} blue spirals host
weak star formation in the outer disks, and have moderately larger disk sizes
than {\em NUV-r} red spirals. Such differences may relate to their environments
and cold gas contents.

\begin{figure}
\plotone{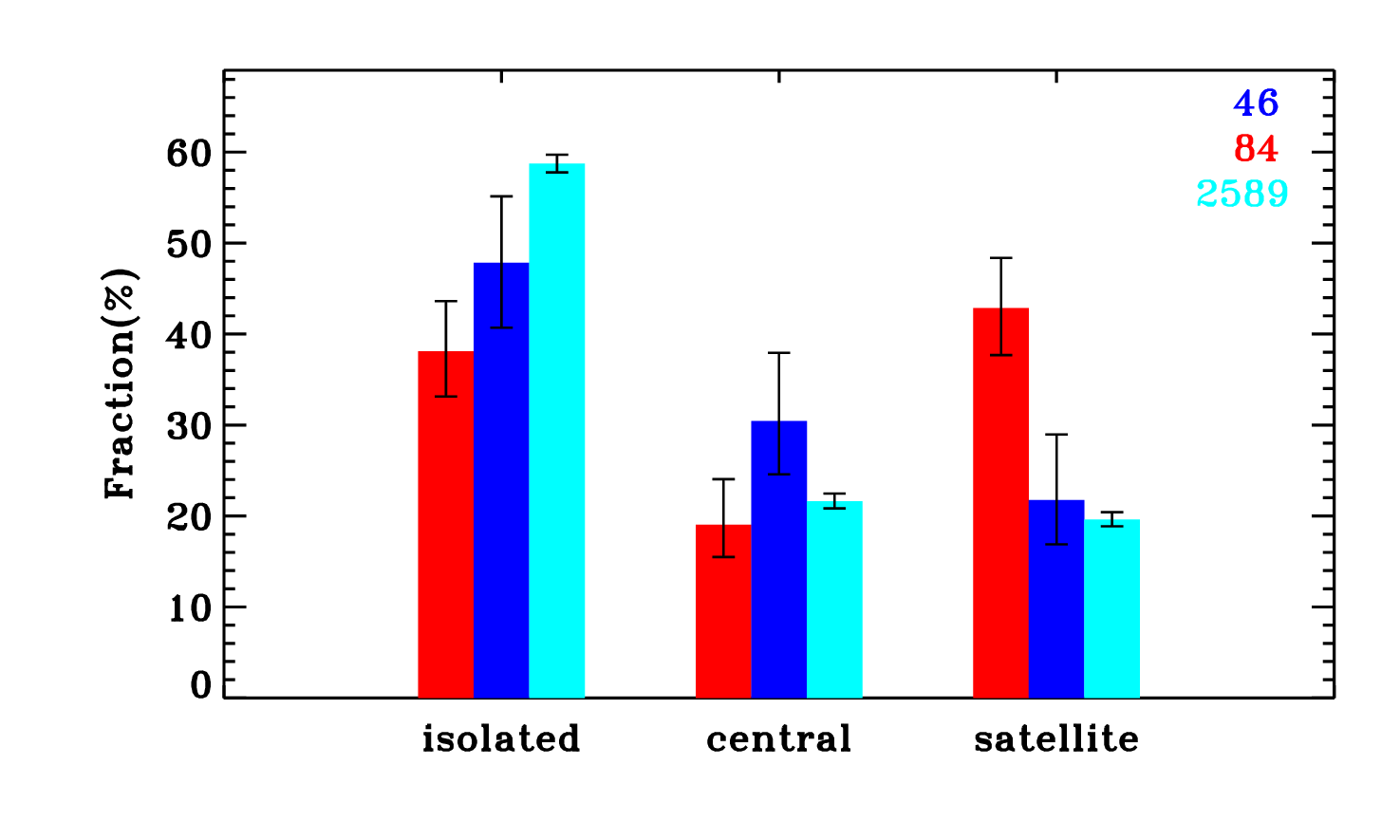}
\caption{
Fractions of {\em NUV-r} blue (blue histogram) and red (red histogram)
spirals, as well as {\em u-r} blue (cyan histogram)
spirals, in different environments.
The error bars represent the 1 $\sigma$ binomial confidence limits, based on the
method of \citet{Cameron2011}.
The number of galaxies in each sample
with environment classifications is shown at the upper-right corner of the
figure.
{\em NUV-r} red spirals show a higher fraction of satellite galaxies than
{\em NUV-r} blue and {\em u-r} blue galaxies, whereas {\em NUV-r} blue spirals
tend to reside in group/cluster centers or isolated environments.
}
\label{environmentstat.eps}
\end{figure}

\begin{figure}
\plotone{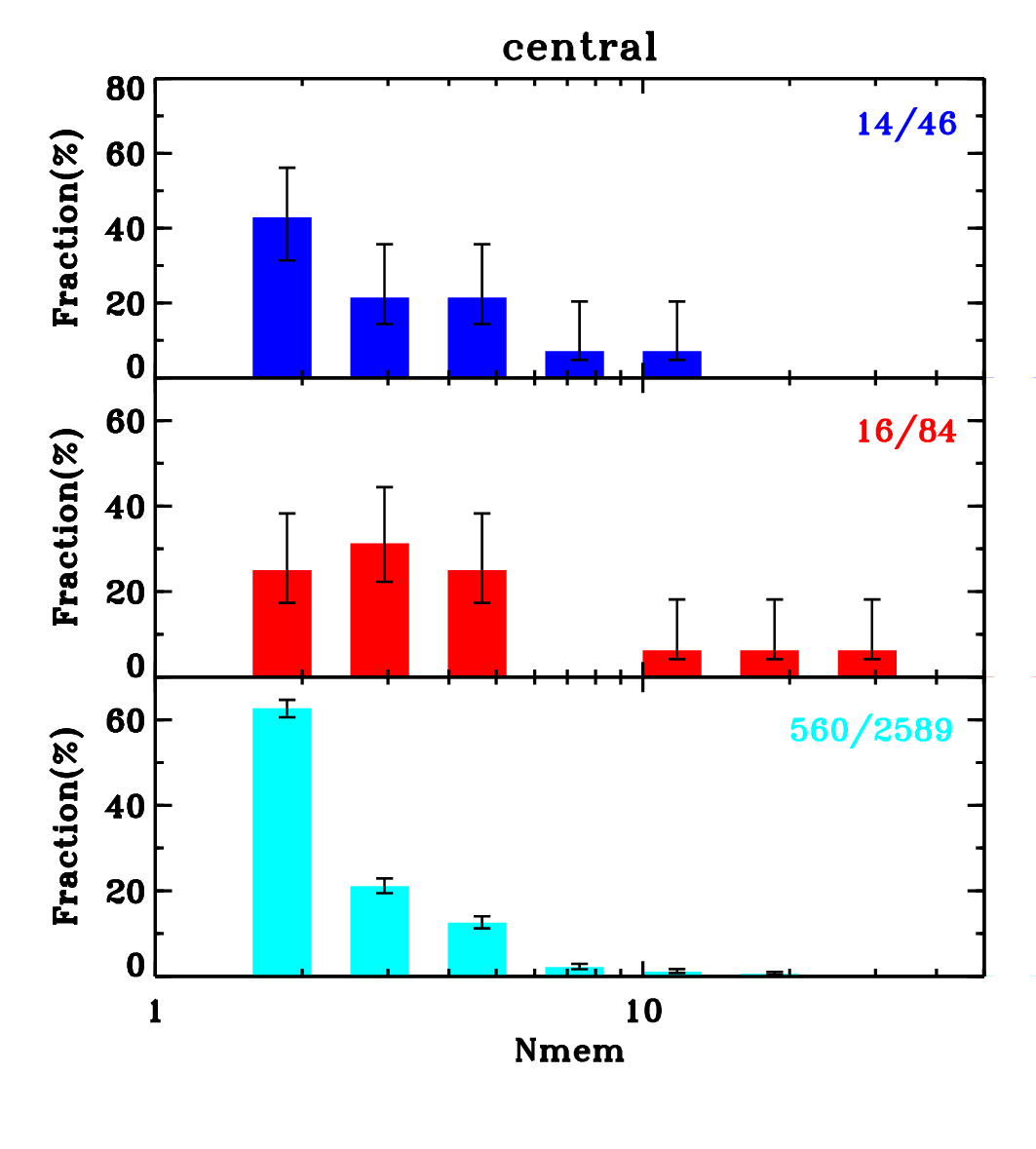}
\caption{
Number of group members for central {\em NUV-r} blue (blue
histogram) and red (red histogram) spirals, as well as {\em u-r} blue (cyan histogram)
spirals.
The error bars represent the 1 $\sigma$ binomial confidence limits, based on the
method of \citet{Cameron2011}.
The number of each sample galaxies is denoted in each panel.
The centrals of all three types of spirals tend to reside in relatively poor groups.
}
\label{Nmem.ps}
\end{figure}

\startlongtable
\begin{deluxetable}{lccc}
\tabletypesize{\footnotesize}
\tablecolumns{4}
\tablewidth{0pt}
\tablecaption{Fractions of {\em NUV-r} blue and red spirals and {\em u-r} blue spirals in different environments}
\tablehead{
\colhead{Sample} & \colhead{Isolated} & \colhead{Central} & \colhead{Satellite}}
\startdata
{\em NUV-r} blue spirals & $48_{-7}^{+7}\%$ & $30_{-6}^{+7}\%$ & $22_{-5}^{+7}\%$ \\
{\em NUV-r} red spirals & $38_{-5}^{+6}\%$ & $19_{-4}^{+5}\%$ & $43_{-5}^{+6}\%$ \\
{\em u-r} blue spirals & $59_{-1}^{+1}\%$ & $22_{-1}^{+1}\%$ & $20_{-1}^{+1}\%$ \\
\enddata
\label{table3}
\end{deluxetable}

Based on the group catalog of \citet{Yang2007}, we characterize the
environments of our sample galaxies. Figure~\ref{environmentstat.eps} compares
the environments of our sample galaxies, along with the comparison sample of
{\em u-r} blue spirals. The corresponding environment fractions are listed
in Table \ref{table3}. It is clear from Figure~\ref{environmentstat.eps} and
Table \ref{table3} that the environments for all three types of spirals are
diverse, but they show different preferences.  Most strikingly, {\em
NUV-r} red spirals exhibit a significantly higher satellite fraction than {\em
NUV-r} blue and {\em u-r} blue spirals. This is not surprising because the
cold circum-galactic gas of spiral galaxies can be removed by ram pressure or
tidal stripping when they enter galaxy clusters or groups, which is a form of
environmental quenching for satellite galaxies \citep[e.g.,][and references
therein]{Gunn1972, Byrd1990, FraserMcKelvie2018, Cortese2021}.  In the
remaining two categories of environments, {\em NUV-r} blue spirals show higher
fractions than their red counterparts. Meanwhile, {\em u-r} blue spirals
display the highest fraction of isolated galaxies among all samples, and a
central fraction comparable to that of {\em NUV-r} red spirals. However, this
result needs to be confirmed by larger samples of {\em NUV-r} blue and red
galaxies selected from optically red spirals. We note that the broad range of
environments occupied by passive spirals were also found in the literature
\citep[e.g.,][]{FraserMcKelvie2018, Mahajan2020}.

We investigate the richness of the host groups or clusters of the central
galaxies in Figure~\ref{Nmem.ps}. It clearly shows that the central galaxies of
all three types of spirals tend to be located in relatively poor groups, with
group sizes typically having fewer than 10 members and mostly fewer than 5. By
examining the DESI images, we find that about 50\% (11/22) of the
isolated {\em NUV-r} blue spiral galaxies show merger or interaction features
in their optical images. This implies that these galaxies, now observed in
isolation, were not truly isolated systems in the past, but rather members of
poor groups. In contrast, almost 60\% of {\em u-r} blue spirals are in
isolation, and they mostly show normal morphologies.

To explore the cold gas content, we cross-matched our samples of {\em NUV-r}
blue and red spirals with an HI-complete sample of {\em u-r} selected red
spirals at $z < 0.05$ from \citet{Guo2020}.  This sample achieves HI
completeness by combining data from the ALFALFA ($\alpha.100$) database
\citep{Giovanelli2005, Haynes2018} and follow-up observations with the
Five-hundred-meter Aperture Spherical radio Telescope (FAST) \citep{Wang2022}.
The cross-match resulted in four {\em NUV-r} blue and twelve {\em NUV-r} red
spirals in the ALFALFA Survey area, and six {\em NUV-r} blue and eight {\em
NUV-r} red spirals observed by FAST. Using a searching radius of 4\arcsec\ , we
found that none of the sample galaxies in the ALFALFA Survey area have HI
detections. For the sample galaxies not in the ALFALFA survey area but observed
by FAST, there are five {\em NUV-r} blue spirals and no {\em NUV-r} red spirals
with a S/N of HI flux greater than 4.7, which is the minimum S/N for the {\em
u-r} selected red spirals detected in the ALFALFA. Therefore, the HI detection
rate for the {\em NUV-r} blue spirals at $z < 0.05$ is about
$50_{-14.4}^{+14.4}\%$ (5/10)\footnote{The $1\sigma$ binomial
uncertainties were calculated using the method of \citet{Cameron2011}.},
comparable to that (58\%) for the {\em u-r} blue spirals reported by
\citet{Guo2020}. The HI mass fractions for these five {\em NUV-r} blue spirals
are mostly less than or around 3\%. They are consistent with the correlation
between {\em NUV-r} color and the HI mass fraction derived by
\citet{Catinella2018} based on 1,179 xGASS galaxies, within the scatters.

For the ten {\em NUV-r} blue spirals covered by the HI observations,
the numbers of isolated, central and satellite galaxies are four, three and
three, respectively. The corresponding numbers of galaxies with HI detections
are two, one, and two. It is hard to draw any conclusion on the relation
between HI acquisition and environment, given the small sample size. More
HI observations are needed to confirm the result.

\section{DISCUSSION}

We have compared the {\em NUV-r} blue and red massive spiral galaxies with red
{\em u-r} colors on the spatial distributions of NUV emission and their
structural properties. The results reveal that {\em NUV-r} red spirals are
fully quenched systems, and they have similar bulge properties to the {\em
NUV-r} blue ones. In comparison, {\em NUV-r} blue spirals exhibit weak star
formation in their outer disks, and possess $\sim 20\%$ larger optical
disks than their {\em NUV-r} red counterparts at fixed disk masses. These
properties suggest that {\em NUV-r} blue spirals may be rejuvenated systems
from quenched galaxies.

The cold gas contents and environments are also consistent with such a
scenario. The {\em NUV-r} blue spirals have an HI detection rate ($\sim 50\%$)
comparable to that of {\em u-r} blue spirals ($\sim 58\%$), but there are no HI
detections in {\em NUV-r} red spirals. Ideally, we might expect a 100\% HI
detection rate for star-forming galaxies. However, the HI data used in the
statistics were mainly drawn from shallow observations, which are biased
towards HI-rich galaxies. Based on ALFALFA observations, \citet{Lin2020} also
found that about 55\% (18/33) ALMaQUEST galaxies were detected in HI, with a
slightly higher S/N than that adopted here ($S/N > 5$). Furthermore, they found
that galaxies with and without HI detections distribute similarly in the
SFMS diagram. The HI mass fractions of our {\em NUV-r} blue spirals are
consistent with the correlation between {\em NUV-r} color and the HI mass
fraction for xGASS galaxies \citep{Catinella2018} within the scatters, but are
lower than those of {\em u-r} blue spirals. This is mainly a selection effect.
The requirement of optically red colors excludes galaxies with significant
rejuvenation, and selects galaxies poorer in HI and star formation.

Regarding the environments, both {\em NUV-r} blue and red spirals occupy a
broad range of environments. This does not contradict the rejuvenation
scenario either. One question that we need to understand is how gas was
accreted to the {\em NUV-r} blue spirals. Therefore, we examine the
morphological features of {\em NUV-r} blue spirals in optical images,
attempting to gain insights into the origin of the HI gas.

\begin{figure}
\plotone{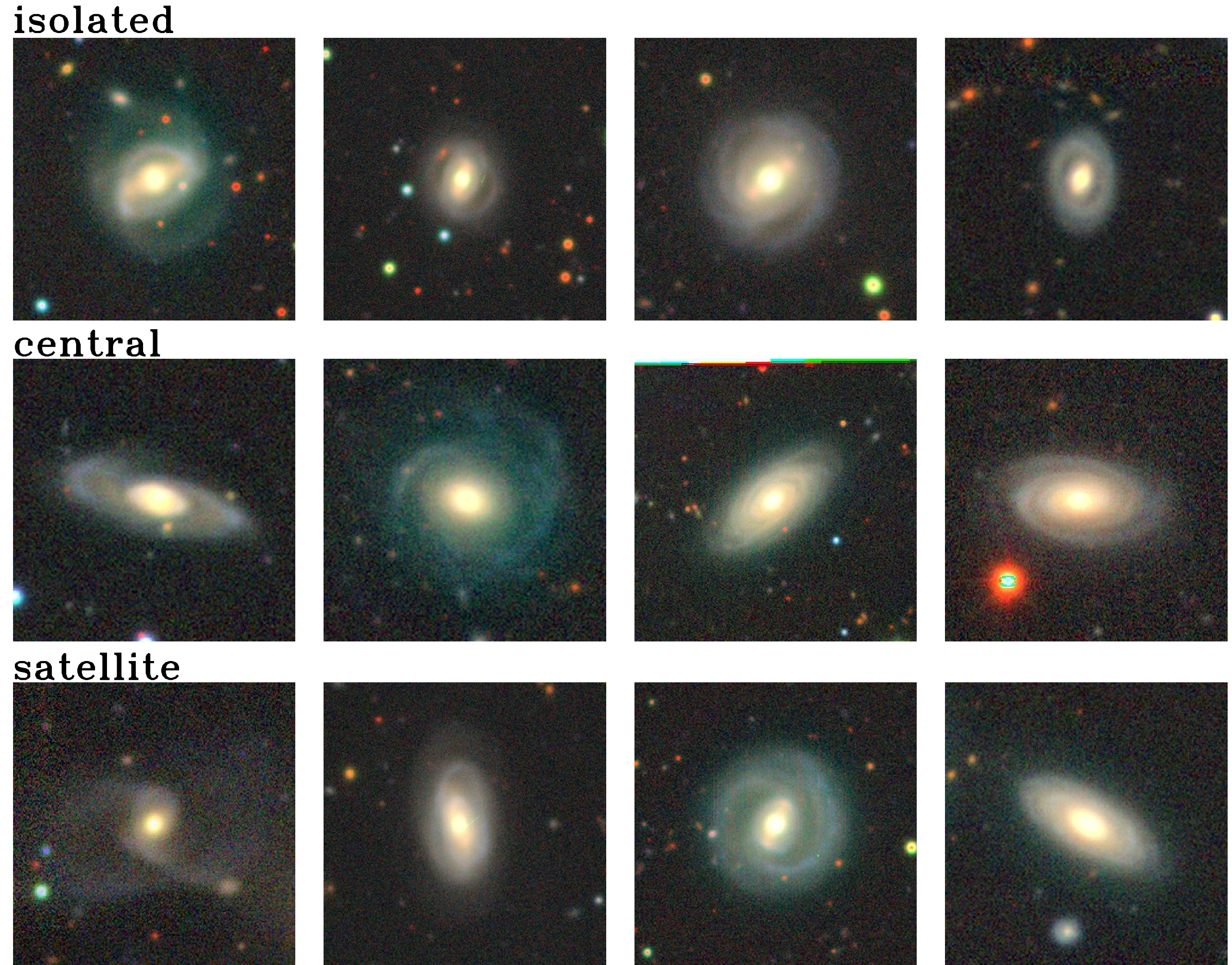}
\caption{
True-color DESI images of example {\em NUV-r} blue spirals in isolated (first row),
central (second row) and satellite (third row) environments.
Across all environments, they exhibit either disturbed (left two panels) or
normal spiral/ring (right two panels) features.
The physical size of each image is 80 $\times$ 80 kpc$^2$.}
\label{Image_blu.ps}
\end{figure}

Across all environments, the {\em NUV-r} blue spirals can be broadly
divided into two morphological categories. One includes galaxies with disturbed
structures, while the other shows normal spiral or ring morphologies, as
illustrated by the examples in Figure~\ref{Image_blu.ps}. This suggests that
{\em NUV-r} blue spirals probably have acquired HI gas via mergers with
gas-rich galaxies or gas accretion from the surroundings. These processes are
closely related to the loss of HI angular momentum. Based on IllustrisTNG-100
simulations, \citet{Lu2022a} revealed that the angular momentum of
circumgalactic medium (CGM) gas, as inherited from the large-scale environment
(CGM spin and orbital motion of neighbouring galaxies) plays a critical role in
regulating star formation and quenching of galaxies.  Particularly,
\citet{Lu2022b} found that although there is not ongoing star formation in the
center of dynamically cold but quenched galaxies any more, new stars are still
forming in the ring-like HI gas structures in their outskirts.

\section{SUMMARY}

Star formation and quenching are fundamental processes governing galaxy
formation and evolution. Galaxy quenching is not always accompanied by
morphological transformation. Optically selected red spirals consist of
quenched galaxies, but also include contamination from galaxies with low
levels of star formation. In order to understand the differences between {\em
NUV-r} blue and red but optically red spirals, we select samples of {\em NUV-r}
blue and red massive ($M_{*} > 10^{10.5} M_{\odot}$) spiral galaxies with
optically ({\em u-r}) red color from the stellar mass catalog of
\citet{Mendel2014} built on SDSS DR7. The {\em NUV-r} blue and red samples
consist of 47 and 86 galaxies, respectively. By comparing the
{\em NUV}, optical and HI properties of the two samples, we find that the {\em
NUV-r} blue but {\em u-r} red spirals are likely rejuvenated systems. Our main
results are summarized as follows.

\begin{enumerate}

\item The locations of {\em NUV-r} blue and red spiral galaxies in the
	SFR vs. stellar mass diagram suggest that {\em NUV-r} red spirals are fully quenched systems,
		whereas {\em NUV-r} blue spirals are below the SFMS
		ridgeline by $\sim 0.5-0.9$ dex, indicating that weak star formation is processing.

\item The optical and {\em NUV} images, as well as the SBPs
	show that the difference between {\em NUV-r} blue and red
		spirals primarily appears in the outer disks (1-3 $R_{\rm e}$), and
		the contrast in the {\em NUV} band is much more distinct than
		that in the optical bands. The {\em NUV-r} color profiles
		further suggest that {\em NUV-r} red spirals have been
		fully quenched from the center to $\sim$ 3 $R_{\rm e}$, which is the
		outermost radius probed in this work. In comparison, {\em
		NUV-r} blue spirals host quenched bulges and inner disks, as
		well as star-forming outer disks, which is the cause for their
		blue {\em NUV-r} colors.
	
\item The mass-size relations show that the bulge components
	of {\em NUV-r} blue and red spirals are similar, and they roughly
		follow the mass-size relation derived for massive ellipticals.
		In contrast, the disk components of {\em NUV-r} blue and red
		spirals distribute differently in the disk mass-size diagram.
		Specifically, {\em NUV-r} red spirals follow the disk relation
		defined by {\em u-r} blue spirals, whereas {\em NUV-r} blue
		spirals lie above this relation, with disks $\sim 1.20$
		times larger than the {\em NUV-r} red spirals at a given disk
		mass. This is a piece of strong evidence for the further disk
		growth via recent star formation activities in the outer disks
		of {\em NUV-r} blue spirals, after the main body formed 6-10
		Gyr ago.      

\item The analyses on environments show that {\em NUV-r} blue and {\em NUV-r}
	red spirals, as well as {\em u-r} blue spirals, reside in all the
		environments investigated in this work. {\em NUV-r} red
		spirals have a higher fraction of satellite galaxies than the
		other two types of blue galaxies, whereas {\em NUV-r} blue
		spirals show a preference for poor group centers or isolated
		environments.

\item Based on a limited subsample with HI observations, we find that {\em
        NUV-r} blue spirals show an HI detection rate of 
                $50_{-14.4}^{+14.4}\%$ (5/10), comparable to that of {\em u-r}
                blue spirals. In contrast, {\em NUV-r} red spirals have no HI
                detections. However, this needs to be verified by a larger
                sample with HI observations.

\end{enumerate}

With the aid of deep optical images from the DESI Legacy Imaging Survey, we
speculate that the moderately larger blue disks of {\em NUV-r} blue but
optically red spirals, relative to their red counterparts, probably formed by
new star formation mainly through interactions or mergers with gas-rich
galaxies or accretion of surrounding HI gas. Therefore, such a sample is
suitable to investigate mild rejuvenation events in the process of galaxy
evolution. We note that current samples are limited, but future high-resolution
UV imaging observations from upcoming telescopes will yield much larger
datasets for such studies.

\begin{acknowledgments}

The authors would like to thank the anonymous referee for the constructive
	comments and suggestions that have improved the paper
	significantly. We also thank Drs. Jing Wang, Cheng Li, Luis Ho, and
	Dandan Xu for helpful discussions. This work is supported by the
	National Key Research and Development Program of China (No.
	2022YFA1602901) and the National Natural Science Foundation of China
	(NSFC, No. 12141301, 12403015).  C.N.Hao also acknowledges the support
	from the science research grants from the China Manned Space Project
	with No.  CMS-CSST-2021-A04, CMS-CSST-2021-A07 and CMS-CSST-2021-B02.
	L.Wang acknowledges the support from the National SKA Program of China
	(No.  2022SKA0110201).
	
	Funding for the creation and distribution of
	the SDSS Archive has been provided by the Alfred P. Sloan Foundation,
	the Participating Institutions, the National Aeronautics and Space
	Administration, the National Science Foundation, the U.S.  Department
	of Energy, the Japanese Monbukagakusho, and the Max Planck Society.
	The SDSS Web site is http://www.sdss.org/.  The SDSS is managed by the
	Astrophysical Research Consortium (ARC) for the Participating
	Institutions. The Participating Institutions are The University of
	Chicago, Fermilab, the Institute for Advanced Study, the Japan
	Participation Group, The Johns Hopkins University, the Korean Scientist
	Group, Los Alamos National Laboratory, the Max-Planck-Institute for
	Astronomy (MPIA), the Max-Planck-Institute for Astrophysics (MPA), New
	Mexico State University, University of Pittsburgh, Princeton
	University, the United States Naval Observatory, and the University of
	Washington.
	
        Some of the data presented in this paper were obtained
        from the Mikulski Archive for Space Telescopes (MAST). STScI is
        operated by the Association of Universities for Research in Astronomy,
        Inc., under NASA contract NAS5-26555. Support for MAST for non-HST data
        is provided by the NASA Office of Space Science via grant NNX09AF08G
        and by other grants and contracts.

	The DESI Legacy Imaging Surveys consist of three individual and
	complementary projects: the Dark Energy Camera Legacy Survey (DECaLS),
	the Beijing-Arizona Sky Survey (BASS), and the Mayall z-band Legacy
	Survey (MzLS). DECaLS, BASS and MzLS together include data obtained,
	respectively, at the Blanco telescope, Cerro Tololo Inter-American
	Observatory, NSF’s NOIRLab; the Bok telescope, Steward Observatory,
	University of Arizona; and the Mayall telescope, Kitt Peak National
	Observatory, NOIRLab. NOIRLab is operated by the Association of
	Universities for Research in Astronomy (AURA) under a cooperative
	agreement with the National Science Foundation.  Pipeline processing
	and analyses of the data were supported by NOIRLab and the Lawrence
	Berkeley National Laboratory (LBNL). Legacy Surveys also uses data
	products from the Near-Earth Object Wide-field Infrared Survey Explorer
	(NEOWISE), a project of the Jet Propulsion Laboratory/California
	Institute of Technology, funded by the National Aeronautics and Space
	Administration. Legacy Surveys was supported by: the Director, Office
	of Science, Office of High Energy Physics of the U.S. Department of
	Energy; the National Energy Research Scientific Computing Center, a DOE
	Office of Science User Facility; the U.S.  National Science Foundation,
	Division of Astronomical Sciences; the National Astronomical
	Observatories of China, the Chinese Academy of Sciences and the Chinese
	National Natural Science Foundation. LBNL is managed by the Regents of
	the University of California under contract to the U.S. Department of
	Energy. The complete acknowledgments can be found at
	https://www.legacysurvey.org/acknowledgment/.

\end{acknowledgments}

\end{document}